\documentclass[12pt]{article}
\usepackage[T1]{fontenc}
\usepackage[utf8]{inputenc} 
\usepackage[english,french,main=english]{babel} 

\usepackage[useregional]{datetime2}

\DeclareUnicodeCharacter{03C6}{$\varphi$}
\DeclareUnicodeCharacter{2082}{\textsubscript{2}}

\usepackage{csquotes} 

\usepackage{biblatex}
\usepackage{imakeidx}
\makeindex

\usepackage{physics}
\usepackage{mathtools} 
\allowdisplaybreaks
\usepackage{amsthm}    
\usepackage{amssymb}   
\usepackage{mathrsfs}  
\usepackage{bbm}       
\usepackage{array}     
\usepackage{upgreek}   

\numberwithin{equation}{section}
\usepackage{appendix}

\usepackage[dvipsnames]{xcolor} 

\usepackage[pdfpagelabels,hypertexnames=false]{hyperref}
\hypersetup{
colorlinks=true,
citecolor=RoyalBlue,
linkcolor=magenta,
urlcolor=Blue
}
\usepackage{orcidlink}

\usepackage{anyfontsize} 
\usepackage[top=2.5cm,bottom=2.5cm,right=2cm,left=2cm]{geometry}
\usepackage{titling}
\usepackage{setspace} 
\usepackage{float} 

\makeatletter
\def\th@plain{%
  \thm@notefont{}%
  \itshape
}
\def\th@definition{%
  \thm@notefont{}%
  \normalfont
}
\def\thmhead@plain#1#2#3{%
  \thmname{#1}\thmnumber{\@ifnotempty{#1}{ }\@upn{#2}}%
  \thmnote{{\the\thm@notefont#3}}%
}
\let\thmhead\thmhead@plain
\makeatother

\newcommand{\mytheoremname}{Theorem}
\newcommand{\mylemmaname}{Lemma}
\newcommand{\mypropname}{Proposition}
\newcommand{\mycorname}{Corollary}
\newcommand{\myexname}{Example}
\newcommand{\myremname}{Remark}
\newcommand{\mydefname}{Definition}

\definecolor{modifcol}{named}{blue}

\definecolor{todocol}{named}{blue}

\newtheorem{theorem}{\protect\mytheoremname}[section]
\newtheorem{lemma}[theorem]{\protect\mylemmaname}
\newtheorem{prop}[theorem]{\protect\mypropname}

\theoremstyle{definition}

\newtheorem{remark}[theorem]{\protect\myremname}

\addto\captionsfrench{%
  \renewcommand{\mytheoremname}{Théorème}%
  \renewcommand{\mylemmaname}{Lemme}%
  \renewcommand{\mypropname}{Proposition}%
  \renewcommand{\mycorname}{Corollaire}%
  \renewcommand{\myexname}{Exemple}%
  \renewcommand{\myremname}{Remarque}%
  \renewcommand{\mydefname}{Définition}%
}
\addto\captionsenglish{%
  \renewcommand{\mytheoremname}{Theorem}%
  \renewcommand{\mylemmaname}{Lemma}%
  \renewcommand{\mypropname}{Proposition}%
  \renewcommand{\mycorname}{Corollary}%
  \renewcommand{\myexname}{Example}%
  \renewcommand{\myremname}{Remark}%
  \renewcommand{\mydefname}{Definition}%
}

\newcounter{hypo}
\newenvironment{hyp}{%
  \begin{enumerate}\setcounter{enumi}{\value{hypo}}\item%
}{%
  \stepcounter{hypo}\end{enumerate}%
}

\newcommand{\egaldef}{\overset{\mathrm{def}}{=}}
\newcommand{\defegal}{\overset{\mathrm{def}}{=}}

\renewcommand{\tilde}{\widetilde}

\renewcommand{\bar}{\overline}
\renewcommand{\leq}{\leqslant}
\renewcommand{\geq}{\geqslant}

\newcommand\N{\mathbb{N}}
\newcommand\R{\mathbb{R}}
\newcommand\C{\mathbb{C}}
\newcommand\eps{\varepsilon}
\newcommand\rmi{\mathrm{i}} 
\newcommand\e{\mathrm{e}} 

\newcommand{\1}{\mathbbm{1}}

\renewcommand\Tr[1]{\mathrm{Tr}\left(#1\right)} 

\addto\captionsenglish{%
}

\addto\captionsenglish{%
}

\newcommand{\Four}{\mathcal{F}} 

\newcommand{\ent}[1]{\left\lfloor #1 \right\rfloor}

\newcommand{\ot}{\otimes}
\newcommand{\ots}{\otimes_{\mathrm{s}}}
\newcommand{\tensn}[2]{\ot^{#2} #1}
\newcommand{\tensymn}[2]{\ots^{#2} #1}

\newcommand{\symformsymb}{\sigma}
\newcommand{\symform}[2]{\symformsymb\left(#1,#2\right)}

\newcommand{\hi}{\mathfrak{h}}
\newcommand{\normhi}[1]{\abs{#1}_\hi}
\newcommand{\conjug}{\mathfrak{c}}

\newcommand{\symop}[1][]{\mathcal{S}_{#1}}

\newcommand{\focks}{\Gamma_{\mathrm{s}}(\hi)}
\newcommand{\focksalg}{\Gamma_{\mathrm{s}}^{\mathrm{fin}}(\hi)}

\newcommand{\wickpol}[1][]{\mathcal{P}_{#1}(\hi)}

\newcommand{\Wick}[2][]{\mathrm{Wick}_{#1}\left(#2\right)}

\newcommand{\scalfield}[1][]{\varphi_{#1}}
\newcommand{\ppde}{P(\scalfield)_2}
\newcommand{\FVclass}[1][]{\mathcal{K}_{#1}(\hi)}

\newcommand{\creasymb}[1][]{\mathrm{a}^{\dagger}_{#1}}
\newcommand{\crea}[2][]{\creasymb[#1]\left(#2\right)}
\newcommand{\annisymb}[1][]{\mathrm{a}_{#1}}
\newcommand{\anni}[2][]{\annisymb[#1]\left(#2\right)}

\newcommand{\weylop}[2][]{\mathrm{W}_{#1}\left(#2\right)}
\newcommand{\nbop}[1][]{\mathrm{N}_{#1}}
\newcommand{\liftop}[1]{\Gamma\left(#1\right)}
\newcommand{\dG}[2][]{\dd \Gamma_{#1}\left(#2\right)}
\newcommand{\gat}[2]{\partial_z^{#1}#2}
\newcommand{\gatbar}[2]{\partial_{\bar{z}}^{#1}#2}
\newcommand{\gatgatbar}[3]{\partial_z^{#1}\partial_{\bar{z}}^{#2}#3}

\newcommand{\smallFVclass}[1][]{\mathfrak{K}_{#1}(\hi)} 
\newcommand{\implsym}{\mathcal{X}(\hi)} 

\makeatletter
\renewcommand{\@fnsymbol}[1]{%
 \ifcase#1\or
 $\,\natural$\or
 $\,\sharp$\or
 $\,\flat$\else
 \@arabic{#1}%
 \fi
}
\makeatother

\title{\textbf{An Egorov-type Theorem for Analytic Wick Observables in Quantum Field Theory}}
\author{Julien \textsc{Malartre}\thanks{Université Sorbonne Paris Nord, Laboratoire Analyse, Géométrie et Applications, LAGA,
CNRS, UMR 7539, F‐93430, Villetaneuse, France. ORCID:\,\orcidlink{0009-0008-9522-2453} 0009-0008-9522-2453. Email: \href{mailto:malartre@math.univ-paris13.fr}{malartre@math.univ-paris13.fr}}}
\date{November 10, 2026}

\begin{document}

\maketitle

\begin{abstract}
    \noindent We provide an exact Egorov-type formula, given by a convergent expansion in the semi-classical parameter $\eps > 0$, for the propagation of a certain class of analytic Wick observables with super-exponentially decaying potentials, under the action of a time-dependent quadratic Hamiltonian. This work generalizes a result established by S.~Breteaux for polynomial symbols, and includes the spatially cutoff $\ppde$ model as well as a variant of the H{\o}egh-Krohn model.
\end{abstract}

\smallskip
\noindent\textbf{Keywords.} Wick observable, analytic symbol, Egorov's theorem, classical limit, $\ppde$ model, H{\o}egh-Krohn model.

\smallskip	
\noindent\textbf{2020 Mathematics Subject Classification.} 81Q20, 81S30, 81T08.

\section{Introduction}\label{sect.intro_noteBr}

The Hepp method \cite{He74}, introduced in 1974 to clarify the crucial role of coherent states in the classical limit, has inspired a wide range of subsequent works on semi-classical dynamics based on coherent states. In particular, it is now well known that coherent states remain localized along classical trajectories, while undergoing a squeezing governed by a unitary Bogoliubov transformation. Such results were first established in finite dimension (see \cite{Ro07} and references therein), and later extended to the mean-field regime \cite{GiVe79}, to infinite-dimensional phase-space analysis \cite{AmNi08}, and to quantum field models \cite{Do81,AmZe14}. The unifying principle of these results is that the control of the remainder term is determined by the quadratic approximation of the Hamiltonian around the classical trajectory. Hence, the asymptotic analysis of the evolution of a Wick observable under a time-dependent quadratic Hamiltonian plays a central role in this context. Such results are provided in \cite[Theorems~3.1 and 3.2]{Br12} for polynomial Wick symbols: while \cite[Theorem~3.1]{Br12} is established through a Dyson expansion which represents the successive terms as time-dependent
integrals, \cite[Theorem~3.2]{Br12} relies on the exact formulas for finite-dimensional Weyl quantization, on the relation between Weyl and Wick symbols, and on a cylindrical approximation argument. These two results are variants of Egorov's theorem (see \cite{Eg69} for the foundational paper and \cite{Le19} for an overview) for polynomial Wick quantization in the sense that, at leading order, quantum conjugation transports observables along the classical flow. The aim of this work is to extend S. Breteaux's result \cite[Theorem~3.2]{Br12} to a certain class of analytic Wick symbols with super-exponentially decaying potentials, including the interactions of the spatially cutoff $\ppde$ model (which is polynomial) and of an analytic variant of the H{\o}egh-Krohn model. Let us emphasize below the main differences between the two settings: 
\begin{enumerate}
    \item The Wick quantization of an analytic symbol is not, in general, relatively bounded with respect to any fixed power of the number operator. Moreover, we will see in Appendix~\ref{app.proof_crucial_estimate_noteBr} that the available bounds on the quantum propagator generated by the quadratic Hamiltonian grow super-exponentially fast in the Hilbert rigging of domains of powers of the number operator: the role of the super-exponential decay assumption \ref{hyp.potentiel_petit_noteBr} is to balance such growth.
    \item Since an analytic symbol involves monomial symbols of arbitrarily high order, the series defining the exponential operator $\e^{\frac{\eps}{2} \Lambda_t}$ on the space of analytic symbols is now infinite. Hence, the definition and properties of this operator require particular attention, see Subsection~\ref{subsect.exp_op_Lambda_t_noteBr}.
    \item Unlike the class of polynomial symbols, the class of analytic symbols under consideration is not necessarily preserved by composition with the classical dynamics: in the analytic case, this operation also transports the complex conjugation, as observed in Subsection~\ref{subsect.compo_classical_flow_noteBr}.
\end{enumerate}

\medskip

\noindent\textbf{Overview of the paper.} In Section~\ref{sect.main_result_noteBr}, after recalling the standard notions of Fock space and polynomial Wick quantization, we introduce the class of analytic symbols under consideration and state our main result, Theorem~\ref{thm.main_thm_noteBr}. We also discuss applications to the spatially cutoff $\ppde$ model and to an analytic variant of the H{\o}egh-Krohn model. Section~\ref{sect.proof_noteBr} is devoted to the proof of Theorem~\ref{thm.main_thm_noteBr}: we define the operator $\Lambda_t$ on analytic symbols and show that the associated exponential operator is well-defined on this class. Then, we study the composition with the classical flow, which transports the complex conjugation, and conclude by an approximation argument. Appendices \ref{app.time_dep_conjugations_noteBr} and \ref{app.proof_crucial_estimate_noteBr} contain the proofs of Propositions \ref{prop.time_dep_conjugations_noteBr} and \ref{prop.inegalite_cruciale_reste_propag_CS_QFT_noteBr}, respectively.

\section{Setting and main result}\label{sect.main_result_noteBr}

In this section, we start by recalling the standard notions related to Fock space and polynomial Wick quantization. Then, we introduce the class of analytic symbols that lies at the heart of this article and for which Wick quantization can be defined rigorously. Our main result, Theorem~\ref{thm.main_thm_noteBr}, describes the propagation of the corresponding Wick observables under a time-dependent quadratic Hamiltonian. Finally, we present two applications: the spatially cutoff $\ppde$ model and a cutoff analytic version of the H{\o}egh-Krohn model.

\subsection{Fock space and Wick quantization}\label{subsect.math_background_noteBr}

Let $\hi$ be a separable complex Hilbert space whose inner product $\braket{\bullet}{\bullet}$ is antilinear with respect to the left variable, with associated norm $\normhi{u} = \sqrt{\braket{u}{u}}$. We endow the complex Hilbert space $\hi$ with a complex conjugation $\conjug : \hi \to \hi$, that is, a norm-preserving, antilinear involution on $\hi$. The bosonic Fock space over $\hi$ is defined as the Hilbert direct sum
\begin{equation}
	 \focks = \displaystyle\bigoplus_{n \in \N} \tensymn{\hi}{n}\,.
\end{equation}
We denote by $\norm{\bullet}_{\focks}$, or simply $\norm{\bullet}$, the Hilbert norm on $\focks$. 
The finite-particle space, defined as the algebraic direct sum 
\begin{equation}\label{focksalg_noteBr}
	 \focksalg = \displaystyle\bigoplus_{n \in \N}^{\mathrm{alg}} \tensymn{\hi}{n}
\end{equation}
is dense in $\focks$. 

\medskip

For any bounded operator $A$ on $\hi$, linear or antilinear, the corresponding lifting operator $\liftop{A}$ is 
\begin{equation}\label{liftop_noteBr}
    \eval{\liftop{A}}_{\tensymn{\hi}{n}} = \tensn{A}{n}\,,
\end{equation}
defined on its maximal domain. Here $\tensn{A}{0}$ is the identity of $\C$ if $A$ is linear, and the usual complex conjugation on $\C$ if $A$ is antilinear. In particular, since $\conjug$ is a complex conjugation on $\hi$, the operator $\liftop{\conjug}$ defines a complex conjugation on the Fock space $\focks$. Furthermore, for all $\lambda \in \C$, we write $\liftop{\lambda}$ for $\liftop{\lambda\, \1}$. We also note that the operator $\liftop{A}$ is self-adjoint when $A$ is linear and self-adjoint.

Let $0 < \eps \leq 1$ be a semi-classical parameter. For a $\C$-linear self-adjoint operator $A$, the second quantization $\dG[\eps]{A}$ is the self-adjoint operator given by:
\begin{equation}\label{def_second_quantization_noteBr}
	   \e^{\frac{\rmi t}{\eps}\, \dG[\eps]{A}} = \liftop{\e^{\rmi t A}}\,.
\end{equation}

\bigskip

Let us now provide minimal material about polynomial Wick quantization. A symbol ${b : \hi \to \C}$ is said to be a monomial symbol if it takes the form 
\begin{equation}
    b(z) = \braket{\tensn{z}{q}}{\tilde{b}(\tensn{z}{p})}_{\tensymn{\hi}{q}} \text{ with } \tilde{b} \in \mathcal{L}(\tensymn{\hi}{p},\tensymn{\hi}{q})
\end{equation}
for some $p,q \in \N$. In this case, we write $b \in \wickpol[p,q]$ and we call $m = p+q$ the order of $b$. We define $\wickpol$ as the space of linear combinations of monomial symbols, called polynomial symbols. The uniqueness of the monomial decomposition ensures that the order of a polynomial symbol is well-defined, as the maximal order of its monomials.
	
	\medskip
	
	For all $b \in \wickpol[p,q]$ and $z,h \in \hi$, we have
	\begin{equation}\label{dév_limité_dérivées_Fréchet_polynôme_Wick_noteBr}
		b(z+h) = \sum_{\substack{0 \leq j \leq q\\0 \leq k \leq p}} \frac{1}{j!} \frac{1}{k!}\, \braket{\tensn{h}{j}}{\gatgatbar{k}{j}{b}(z) (\tensn{h}{k})}\,,
	\end{equation}
	where 
	\begin{multline}\label{def_gatgatbar_noteBr}
		\gatgatbar{k}{j}{b}(z) = \frac{p!}{(p-k)!}\, \frac{q!}{(q-j)!}\, \big(\bra{\tensn{z}{q-j}} \ot (\tensn{\1}{j})\big)\,\tilde{b}\, \symop[p]\big(\ket{\tensn{z}{p-k}} \ot (\tensn{\1}{k})\big)\\ \in \mathcal{L}(\tensymn{\hi}{k},\tensymn{\hi}{j})\,.
	\end{multline}
	Here, $\symop[n]$ is the orthogonal projection of $\tensn{\hi}{n}$ onto $\tensymn{\hi}{n}$. Formulas \eqref{dév_limité_dérivées_Fréchet_polynôme_Wick_noteBr} and \eqref{def_gatgatbar_noteBr} provide a notion of Gâteaux differentials for polynomial symbols.
	
	\medskip
	
	For fixed $0 < \eps \leq 1$, the semi-classical Wick quantization of a monomial symbol $b \in \wickpol[p,q]$ is defined on $\focksalg$ by 
	\begin{equation}\label{quantif_Wick_poly_tensym_noteBr}
		\eval{\Wick[\eps]{b}}_{\tensymn{\hi}{n}} = \1_{[p,+\infty)}(n)\, \frac{\sqrt{n!\,(n+q-p)!}}{(n-p)!}\, \eps^{\frac{p+q}{2}}\, \symop[n+q-p]\left(\tilde{b} \ot (\tensn{\1}{n-p})\right)\,.
	\end{equation}
	The expression $\Wick[\eps]{b}$ is extended by linearity to any polynomial symbol $b \in \wickpol$, and $b$ is then called the Wick symbol of the operator $\Wick[\eps]{b}$. The inclusion $\Wick[\eps]{\bar{b}} \subset \Wick[\eps]{b}^{\dagger}$, where the notation $\dagger$ stands for the adjoint operator and where $\bar{b}(z)$ is the complex conjugate of $b(z)$, shows that the operators $\Wick[\eps]{b}$ are closable. We keep the same notation for their closure.
	
	\medskip
	
	For further discussion about polynomial Wick quantization, we refer to \cite{AmNi08} and \cite{DeGe00} (see also \cite{Fo89}).

\subsection{Evolution of analytic Wick observables}\label{subsect.evolution_analytic_symbols_noteBr}

Let us now define the class $\FVclass$ of symbols $F : \hi \to \C$ of the form 
\begin{equation}\label{symb_Fvclass_noteBr}
    F_V(z) = \sum_{j = 0}^\infty \braket{\frac{\tensn{(z+\conjug z)}{j}}{\sqrt{j!}}}{V^{(j)}}\,,
\end{equation}
where $V = \Big(V^{(j)}\Big)_{j \in \N}$ is an element of $\focks$, called the potential associated with $F_V \in \FVclass$. Note that the series \eqref{symb_Fvclass_noteBr} is absolutely convergent for fixed $z \in \hi$, with the estimate \[\abs{F_V(z)} \leq \e^{\normhi{z+\conjug z}^2/2}\, \norm{V}_{\focks}\,.\]

Endowed with the inner product 
\[
\braket{F_{V_1}}{F_{V_2}}_{\FVclass} = \braket{V_1}{V_2}_{\focks}\,,
\] 
which is well-defined by the injectivity of $V \mapsto F_V$, the space $\FVclass$ is a Hilbert space which is isometrically isomorphic to $\focks$. 

\medskip

We note that there is no inclusion relation between the space $\FVclass$ and the space of polynomial symbols $\wickpol$: for instance, we have $\normhi{z}^2 \in \wickpol \setminus \FVclass$ since it is not a function of $z + \conjug z$, and, for $V = \left(\displaystyle\frac{\tensn{u}{j}}{\sqrt{j!}}\right)_{j \in \N}$ where $u \in \hi \setminus \{0\}$ satisfies $\conjug u = u$, we have \[F_V(z) = \exp\left(2\, \Re\braket{z}{u}\right) \in \FVclass \setminus \wickpol\,.\]

\bigskip
	
It turns out that the usual polynomial Wick quantization can be naturally extended to the class $\FVclass$: for more details, we refer the reader to \cite[Subsection~3.2]{AmZe14}. The semi-classical Wick quantization of an analytic symbol $F_V \in \FVclass$ is denoted by $\Wick[\eps]{F_V}$, and coincides with the closure of the operator defined by \eqref{quantif_Wick_poly_tensym_noteBr} whenever $F_V \in \wickpol \cap \FVclass$.

\medskip
	
Inspired by the work \cite{AmZe14} on the propagation of coherent states, we consider symbols ${F_V \in \FVclass}$ whose potential $V$ has a super-exponential decay rate. This assumption will be discussed in Remark~\ref{rem.on_main_thm_noteBr} below. More precisely, let us denote by $\smallFVclass$ the class of symbols $F_V \in \FVclass$ with potential $V$ fulfilling the following condition: 
\begin{hyp}\label{hyp.potentiel_petit_noteBr}
	$V \in D\Big(\e^{\gamma\, \Gamma(\lambda)}\Big) \,$ for some $\gamma > 0$ and $\lambda > 1$, that is, $\,\displaystyle\sum_{j=0}^{\infty} \e^{2 \gamma \lambda^j}\, \norm{V^{(j)}}_{\tensn{\hi}{j}}^2 < + \infty$.
\end{hyp}
We recall that $\liftop{\lambda}$ denotes the lifting operator corresponding to $\lambda\, \1$, see \eqref{liftop_noteBr}.

\medskip

The aim of this work is to study the propagation of the Wick observable $\Wick[\eps]{F_V}$ with $F_V \in \smallFVclass$, under the dynamics governed by a quadratic symbol of the form 
\begin{equation}\label{def_quadratic_symbol_noteBr}
    Q_t(z) = \braket{z}{\mathfrak{a}_t\, z} + \Im\braket{\mathfrak{b}_t}{\tensn{z}{2}}\,.
\end{equation}
More precisely, following \cite{Br12}, we assume that ${\mathfrak{b}_t \in C^0(\R,\tensymn{\hi}{2})}$ and that $(\mathfrak{a}_t)_{t \in \R}$ is a family of self-adjoint operators on $\hi$ satisfying the assumptions of \cite[Theorem~X.70]{ReSi75} on every compact interval: 
\begin{enumerate}\label{hyp_X.70_reed_simon}
    \item The operators $\mathfrak{a}_t$ have a common domain, from which it follows by the closed graph theorem that each operator $T(t,s) = (\mathfrak{a}_t - \rmi)\, (\mathfrak{a}_s - \rmi)^{-1}-\1$ is bounded,
    \item For each $z \in \hi$, the map $(t,s) \mapsto (t-s)^{-1}\, T(t,s)\, z$ is uniformly strongly continuous and uniformly bounded for $t \neq s$ lying in any fixed compact interval,
    \item For each $z \in \hi$, $T(t)\, z = \underset{s \nearrow t}{\lim}\, (t - s)^{-1}\, T(t,s)\, z$ exists uniformly for $t$ in every compact interval, and $T(t)$ is bounded and strongly continuous in $t$.
\end{enumerate}
Under these assumptions, the family of operators $(\mathfrak{a}_t)_{t \in \R}$ generates a unitary propagator $\big(u_\mathfrak{a}(t)\big)_{t \in \R}$ satisfying 
\begin{equation}\label{equation_alphat_noteBr}
    \left\{\begin{array}{l}
        \rmi\, \partial_t\, u_\mathfrak{a}(t) = \mathfrak{a}_t\, u_\mathfrak{a}(t)\\
    u_\mathfrak{a}(0) = \1\,.
    \end{array}
    \right.
\end{equation}
Furthermore, the classical dynamics $\upphi_t$ associated with the differential equation 
\begin{equation}\label{classical_evolution_noteBr}
	\left\{\begin{array}{l}
		\rmi\, \dot{z}_t = \gatbar{}{Q_t}(z_t)\\
		\eval{z_t}_{t=0} = z_0 \in \hi
		\end{array}\right.
\end{equation}
is well-defined, see \cite[Section~4]{Br12}. Each map ${\upphi_t : \hi \to \hi}$ is a symplectomorphism, that is, an $\R$-linear bijection which preserves the canonical symplectic form $\symform{z_1}{z_2} = \Im\braket{z_1}{z_2}$. In fact, each map $\upphi_t$ is bounded, and the classical dynamics $\upphi_t$ is well-defined for all $t \in \R$ (see Subsection~\ref{subsect.op_Lambda_t_noteBr}, Steps 1 and 2).

On the other hand, for any semi-classical parameter $0 < \eps \leq 1$, the quantum dynamics associated with $Q_t$ is given by the linear evolution equation 
\begin{equation}\label{quantum_evolution_noteBr}
		\left\{\begin{array}{l}
			\rmi \eps\, \partial_t \psi_t = \Wick[\eps]{Q_t}\, \psi_t\\
			\eval{\psi_t}_{t=0} = \psi_0\,,
		\end{array}\right.
\end{equation} 
where $\Wick[\eps]{\braket{z}{\mathfrak{a}_t\, z}} = \dG[\eps]{\mathfrak{a}_t}$. In what follows, we denote by 
\begin{equation}\label{nbop1_noteBr}
    \nbop : \psi = \left(\psi^{(n)}\right)_{n \in \N} \in \focks \mapsto \left(n\, \psi^{(n)}\right)_{n \in \N}
\end{equation}
the number operator, which is $\eps$-independent. When formulated in an appropriate Hilbert rigging involving domains of powers of $\nbop^{1/2}$, the evolution \eqref{quantum_evolution_noteBr} generates a unitary propagator $(U_t)_{t \in \R}$, see \cite[Section~5]{Br12} and Appendix~\ref{app.proof_crucial_estimate_noteBr}. We note that $U_t$ does not depend on $\eps$, since $\Wick[\eps]{Q_t} = \eps\, \Wick[1]{Q_t}$.

\medskip

We now state the main result of this article, providing an Egorov-type formula in the semi-classical parameter $\eps > 0$ for the propagation of Wick observables with symbol in $\smallFVclass$ under the action of a time-dependent quadratic Hamiltonian. As we will see in Section~\ref{sect.proof_noteBr}, the proof of Theorem~\ref{thm.main_thm_noteBr} (see below) relies heavily on the result \cite[Theorem~3.2]{Br12} established by S.~Breteaux for polynomial symbols, and the underlying geometric mechanism at stake is very similar to the one introduced in \cite{Br12}. In particular, both results \cite[Theorem~3.2]{Br12} and Theorem~\ref{thm.main_thm_noteBr} involve a time-dependent operator of the form 
\begin{equation}\label{op_Lambda_t_noteBr}
    \Lambda_t b(z)
    = -2\, \Tr{A_t^{\dagger} A_t\, \gatgatbar{}{}{b}(z)}
    + \braket{v_t}{\gatbar{2}{b}(z)}_{\tensn{\hi}{2}}
    + \braket{\gat{2}{b}(z)}{v_t}_{(\tensn{\hi}{2})^*,\tensn{\hi}{2}}\,. 
\end{equation}
Here, 
\begin{itemize}
    \item the Gâteaux differentials $\gat{2}{b}(z)$, $\gatbar{2}{b}(z)$ and $\gatgatbar{}{}{b}(z)$ are given by \eqref{def_gatgatbar_noteBr},
    \item $\upphi_t = L_t + A_t$ is the decomposition of $\upphi_t$ into its $\C$-linear and $\C$-antilinear parts,
    \item the adjoint of a $\C$-antilinear operator $A$ is the unique $\C$-antilinear operator $A^\dagger$ satisfying $\braket{z_1}{A\, z_2} = \braket{z_2}{A^\dagger\, z_1}$ for all $z_1,z_2 \in \hi$ (see \cite[Definition~A.3]{Br12}),
    \item $v_t \in \tensn{\hi}{2}$ is the unique vector satisfying $\braket{z_1 \ot z_2}{v_t} = \braket{z_1}{L_t^{\dagger}\, A_t\, z_2}$ for all ${z_1,z_2 \in \hi}$.
\end{itemize}
The rigorous definition and properties of the operator $\Lambda_t$ will be discussed in Section~\ref{sect.proof_noteBr}.

	\begin{theorem}\label{thm.main_thm_noteBr}
    Let $V \in \focks$ satisfy Assumption~\ref{hyp.potentiel_petit_noteBr}. Then, for all $t_0 > 0$, there exists $\eps_*(t_0) > 0$ such that, for all $t \in [-t_0,t_0]$, the identity 
    \begin{equation}\label{generalisation_formule_Breteaux_conj_Wick_par_propag_quad}
        U_t^{\dagger}\, \Wick[\eps]{F_V}\, U_t = \Wick[\eps]{\e^{\frac{\eps}{2}\, \Lambda_t} (F_V \circ \upphi_t)}
    \end{equation}
    holds for all $0 < \eps \leq \eps_*(t_0)$ in the space $\mathcal{L}\left(D\Big(\liftop{\sqrt{3}}\Big),\focks\right)$. Here,
    \begin{itemize}
        \item $\Wick[\eps]{F_V}$ is the semi-classical Wick quantization of $F_V$,
        \item $\Lambda_t$ denotes the operator defined on $\wickpol$ by \eqref{op_Lambda_t_noteBr} and extended to analytic symbols in Subsections \ref{subsect.op_Lambda_t_noteBr}, \ref{subsect.exp_op_Lambda_t_noteBr} and \ref{subsect.compo_classical_flow_noteBr},
        \item the right-hand side is the Wick quantization of a symbol in a time-dependent analytic class $\FVclass[t]$, see Remark~\ref{rem.on_main_thm_noteBr}.
    \end{itemize}
\end{theorem}

\begin{remark}\label{rem.on_main_thm_noteBr}
    Let us make a few comments on Theorem \ref{thm.main_thm_noteBr}:
    \begin{enumerate}
        \item The leading term on the right-hand side of \eqref{generalisation_formule_Breteaux_conj_Wick_par_propag_quad} is equal to $\Wick[\eps]{F_V \circ \upphi_t}$: hence, up to semi-classical corrections, the quantum dynamics transports observables in the class $\smallFVclass$ along the classical flow. In this sense, formula \eqref{generalisation_formule_Breteaux_conj_Wick_par_propag_quad} provides a variant of Egorov's theorem in the framework of Wick quantization, with explicit semi-classical corrections. We refer the reader to \cite{Le19} and the references mentioned therein for an overview of Egorov-type results.
        \item The symbol $\e^{\frac{\eps}{2} \Lambda_t}(F_V \circ \upphi_t)$ of the right-hand side does not necessarily belong to the class $\FVclass$. However, we will see that this symbol belongs to an analytic class $\FVclass[t]$, defined like $\FVclass$ but with $\conjug$ replaced by a time-dependent conjugation $\conjug_t$; Wick quantization is still well-defined on $\FVclass[t]$. We refer to Subsection~\ref{subsect.compo_classical_flow_noteBr} for further discussion.
        \item The super-exponential decay assumption \ref{hyp.potentiel_petit_noteBr}, although possibly suboptimal, arises naturally when one wishes to estimate a quantity of the form $\norm{\Wick[\eps]{F_V}\, U_t\, \psi}$ for $F_V \in \smallFVclass$ and ${\psi \in \focksalg}$, see Proposition~\ref{prop.inegalite_cruciale_reste_propag_CS_QFT_noteBr}. In short, the role of this assumption is to balance the growth of the norms $\norm{U_t}_{\mathcal{L}(D(\nbop^{k/2}))}$, whose upper bounds obtained by standard techniques grow super-exponentially fast in $k$ (see Proposition~\ref{prop.prop_4.5_AmZe_noteBr}).
        \item The result presented in Theorem~\ref{thm.main_thm_noteBr} is useful to study a specific family of abstract evolution problems in the interaction representation. More precisely, let us consider an evolution problem of the form  
	   \begin{equation}\label{complicated_equation_noteBr}
	       \left\{
	       \begin{array}{l}
	            \rmi \eps\, \partial_t \psi_t = \left(\Wick[\eps]{Q_t} + \Wick[\eps]{F_V}\right)\, \psi_t\,\\
	            \eval{\psi_t}_{t=0} = \psi_0\,,
	       \end{array}
	       \right.
	   \end{equation}
	   where $Q_t$ is the quadratic symbol given by \eqref{def_quadratic_symbol_noteBr} and $F_V \in \smallFVclass$ with ${\liftop{\conjug}V = V}$. Here, $\liftop{\conjug}$ is the lifting operator corresponding to the complex conjugation $\conjug$, see \eqref{liftop_noteBr}. In the interaction representation, writing $\psi_t = U_t\, \tilde{\psi}_t$ with $U_t$ the unitary propagator of \eqref{quantum_evolution_noteBr} and using formula \eqref{generalisation_formule_Breteaux_conj_Wick_par_propag_quad}, the equation above becomes 
	   \begin{equation}\label{complicated_equation_transformed_noteBr}
	       \left\{
	       \begin{array}{l}
	            \rmi \eps\, \partial_t \tilde{\psi}_t = \Wick[\eps]{\e^{\frac{\eps}{2} \Lambda_t} (F_V \circ \upphi_t)}\, \tilde{\psi}_t\,\\
	            \eval{\tilde{\psi}_t}_{t=0} = \psi_0
	       \end{array}
	       \right.
	   \end{equation}
	  for $\abs{t} \leq t_0$ and $0 < \eps \leq \eps_*(t_0)$. As noted in the second item of this remark, the symbol \[\e^{\frac{\eps}{2} \Lambda_t} (F_V \circ \upphi_t) \egaldef F_{V_{t,\eps}}^{\conjug_t}\] belongs to a certain analytic class of symbols $\FVclass[t]$, involving a time-dependent complex conjugation $\conjug_t$, and the associated potential $V_{t,\eps}$ satisfies explicit estimates involving $V$, see Subsection~\ref{subsect.compo_classical_flow_noteBr}. Hence, the initial problem \eqref{complicated_equation_noteBr} reduces to the Cauchy problem  
	   \begin{equation}
	       \left\{
	       \begin{array}{l}
	            \rmi \eps\, \partial_t \tilde{\psi}_t = \Wick[\eps]{F_{V_{t,\eps}}^{\conjug_t}}\, \tilde{\psi}_t\,\\
	            \eval{\tilde{\psi}_t}_{t=0} = \psi_0\,.
	       \end{array}
	       \right.
	   \end{equation}
	   For $\psi_0 \in D\Big(\liftop{\sqrt{3}}\Big)$, this problem can be studied in the Hilbert rigging 
	   \[
	   D\Big(\liftop{\sqrt{3}}\Big) \subset \focks \subset D\Big(\liftop{\sqrt{3}}\Big)^*\,,
	   \]
	   using abstract results such as \cite[Corollary C.4]{AmBr12}, after checking some symmetry and time-regularity assumptions for $\Wick[\eps]{F_{V_{t,\eps}}^{\conjug_t}}$, as well as commutator bounds involving the operator $\liftop{\sqrt{3}} = 3^{\nbop/2}$.
    \end{enumerate}
\end{remark}

\subsection{Applications to quantum field models}

In this subsection, we present two applications of Theorem~\ref{thm.main_thm_noteBr} to landmark examples of constructive quantum field theory: the spatially cutoff $\ppde$ model and an analytic variant of the H{\o}egh-Krohn model. In both cases, we will see that the quadratic symbol is time-independent and takes the form $Q_t(z) = \braket{z}{\omega\, z}$, where $\omega$ is the dispersion relation. Hence, we have $\mathfrak{b}_t = 0$ in \eqref{def_quadratic_symbol_noteBr} and the classical flow $\upphi_t = \e^{-\rmi t \omega}$ is unitary, so the operator $\Lambda_t$ given by \eqref{op_Lambda_t_noteBr} is identically zero. Consequently, formula \eqref{generalisation_formule_Breteaux_conj_Wick_par_propag_quad} simply reads
\begin{equation}
    U_t^{\dagger}\, \Wick[\eps]{F_V}\, U_t = \Wick[\eps]{F_V \circ \upphi_t} 
\end{equation}
for all $t \in \R$ and $\eps > 0$ sufficiently small, in $\mathcal{L}\left(D\left(\liftop{\sqrt{3}}\right),\focks\right)$.

\subsubsection*{The spatially cutoff $\ppde$ model}

A particular case of a free Bose field Hamiltonian with an interaction term of the form $\Wick[\eps]{F_V}$ is provided by the spatially cutoff $\ppde$ Hamiltonian, which corresponds to a polynomial interaction in two-dimensional space-time. Although this interaction falls within the framework of polynomial symbols considered in \cite{Br12}, we include this model because of its central role in constructive quantum field theory, and because the analytic-potential formalism offers a different viewpoint.

\medskip

We follow the construction presented in \cite[Section~6.1]{DeGe00}, and we refer to this work for a more detailed study of the $\ppde$ model. We work on the Hilbert space $\hi = L^2(\R,\dd{k})$ endowed with the complex conjugation $\conjug z(k) = \bar{z(-k)}$. For $m_0 > 0$ fixed, we define the dispersion relation
	\begin{equation}\label{dispersion_relation_noteBr}
	    \omega(k) = \left(m_0^2 + k^2\right)^{1/2}\,,
	\end{equation}
and we define: 
\begin{itemize}
    \item $P = \displaystyle\sum_{j = 0}^{2n} a_j\, X^j$, a real polynomial of degree $2n$ with $a_{2n} > 0$, so that $P$ is bounded below,
    \item $g \in L^1(\R) \cap L^2(\R)$, a non-negative cutoff function.
\end{itemize}
Let us introduce the kernels
\begin{equation}\label{def_kernel_wj_noteBr}
    w_j(k_1,\dots,k_j) = \frac{\Four g(k_1 + \cdots + k_j)}{\sqrt{\omega(k_1) \cdots \omega(k_j)}}\,, 
\end{equation}
where $\Four g(k) = \displaystyle\int_{\R} g(x)\, \e^{-\rmi\, k\, x} \dd{x}$ denotes the Fourier transform of $g$, and let $w_0 = \norm{g}_1$. Then, we have $w_j \in L^2_{\rm sym}(\R^j) \simeq \tensymn{L^2(\R)}{j}$ for all $j \in \N^*$, see \cite[Lemma~6.1]{DeGe00}.  For fixed $0 < \eps \leq 1$, the $\eps$-dependent spatially cutoff $\ppde$ Hamiltonian is given by
\begin{equation}
    H_\eps = \dG[\eps]{\omega} + \Wick[\eps]{F_V}\,, \text{ where } F_V(z) = \sum_{j = 0}^{2n} \braket{\frac{\tensn{(z+\conjug z)}{j}}{\sqrt{j!}}}{\sqrt{j!}\, a_j\, w_j}\,.
\end{equation}
Here, $\dG[\eps]{\omega}$ denotes the second quantization of $\omega$, see \eqref{def_second_quantization_noteBr}.

For the $\ppde$ model, we have $F_V \in \wickpol \cap \FVclass$, so Assumption~\ref{hyp.potentiel_petit_noteBr} is automatically satisfied.	

\medskip

\subsubsection*{The analytic H{\o}egh-Krohn model}

Another example of a free Bose field Hamiltonian with an analytic interaction is given by a modified version of the H{\o}egh-Krohn model. Although this model is less directly tied to particle physics than $\ppde$, it is historically significant as one of the first known models with a non-polynomial interaction satisfying the Wightman axioms (see \cite{HK71bis,AlHK74}). Here, we present a modified version of the H{\o}egh-Krohn model, introduced in \cite[Subsection~3.4]{AmZe14}, incorporating spatial and momentum cutoffs and focusing on \enquote{analytic} perturbations of the field operator.

\medskip

On $\hi = L^2(\R^d,\dd{k})$ endowed with the complex conjugation $\conjug z(k) = \bar{z(-k)}$ and for fixed $m_0 > 0$, let $\omega(k) = (m_0^2+\abs{k}^2)^{1/2}$ be the dispersion relation. For all $\kappa \geq 1$, let us define
	\begin{equation}
		u_\kappa(k) = \omega(k)^{-1/2}\, \Four\chi\left(\frac{k}{\kappa}\right)\,,
	\end{equation}
	where $\chi \in C_c^\infty(\R^d)$, $\chi \geq 0$ is radial, of total mass 1 and supported in the unit ball. Here, $\Four\chi(k) = \displaystyle\int_{\R^d} \chi(x)\, \e^{-\rmi k \cdot x}\, \dd{x}$ denotes the Fourier transform of $\chi$. Let $G(y) = \displaystyle\sum_{j \in \N} a_j\, y^j$, $a_j \in \R$, be an entire function. For $r \geq 1$ fixed, the spatially cutoff H{\o}egh-Krohn model associated with $G$ corresponds to the Hamiltonian 
	\begin{equation}\label{def_spatially_cutoff_HK_Hamiltonian_noteBr}
	    H^{\mathrm{HK}}_\eps = \dG[\eps]{\omega} + \Wick[\eps]{F_V}\,,
	\end{equation}
	where 
	\begin{equation*}
	\left\{
	    \begin{array}{l}
	        V^{(j)} = \sqrt{j!}\, a_j\, f_j\,,\quad \forall\, j \in \N^*\,,\\
	        \text{ with } f_j(k_1,\dots,k_j) = \displaystyle\prod_{m=1}^j u_\kappa(k_m)\, \Four\left(\1_{\abs{x} \leq r}\right)(k_1 + \cdots + k_j) \in L^2_{\mathrm{sym}}(\R^{dj}) \simeq \tensymn{L^2(\R^d)}{j}\,,\\
	        {} \\
	        V^{(0)} = a_0\, {\rm Vol}_d(r)\,,
	    \end{array}
	    \right.
	\end{equation*}
	where ${\rm Vol}_d(r)$ denotes the volume of the Euclidean ball of radius $r$ in $\R^d$. For all $j \in \N^*$, we have 
	\begin{equation*}
	    \norm{V^{(j)}}_{\tensn{\hi}{j}}^2 \leq j!\, a_j^2\, \norm{u_\kappa}_{2}^{2j}\, \norm{\Four \left(\1_{\abs{x} \leq r}\right)}_\infty^2 \leq j!\, a_j^2\, \left(\frac{\kappa^d}{m_0}\, \norm{\Four \chi}_2^2\right)^j\,{\rm Vol}_d(r)^2\,.
	\end{equation*}
	Hence, if the coefficients $a_j$ satisfy  
	   \begin{equation}
	        \sum_{j=0}^\infty \e^{2 \gamma (\lambda')^j} \, j! \, a_j^2 < +\infty \quad \text{ for some $\gamma >0$ and $\lambda' > 1$\,,}
	   \end{equation}
	   Assumption~\ref{hyp.potentiel_petit_noteBr} holds for $V$ for any $\gamma' \in (0,\gamma)$ and the same $\lambda' > 1$. For instance, one can pick $a_j = \e^{-\lambda^j}$ where $\lambda > \lambda' > 1$.
	   
\section{Proof of the main result}\label{sect.proof_noteBr}

This section is devoted to the proof of our main result, Theorem~\ref{thm.main_thm_noteBr}. After introducing the notion of implementability in Subsection~\ref{subsect.implementability_noteBr}, we provide a precise definition of the operator $\Lambda_t$ given by \eqref{op_Lambda_t_noteBr} and we show that $\Lambda_t$ is well-defined pointwise on $\FVclass$, see Subsection~\ref{subsect.op_Lambda_t_noteBr}. Then, in Subsection~\ref{subsect.exp_op_Lambda_t_noteBr}, we prove that $\Lambda_t$ preserves $\smallFVclass$ and that the exponential operator $\e^{\frac{\eps}{2}\, \Lambda_t}$ is well-defined on $\smallFVclass$ for $0 < \eps \leq \eps(t)$ sufficiently small, by establishing useful norm estimates. In Subsection~\ref{subsect.compo_classical_flow_noteBr}, we prove that composition with the classical flow $\upphi_t$ sends the class $\smallFVclass$ into a new time-dependent class $\FVclass[t]$ with time-dependent complex conjugation, on which the operator $\e^{\frac{\eps}{2}\, \Lambda_t}$ is also well-defined and satisfies explicit estimates. Finally, in Subsection~\ref{subsect.end_proof_noteBr}, we conclude the proof of Theorem~\ref{thm.main_thm_noteBr}, first for polynomial truncations, then for general analytic symbols by passing to the limit and using density arguments.

\subsection{Implementable symplectomorphisms}\label{subsect.implementability_noteBr}
	   
	   In order to define the operator $\Lambda_t$ appearing in Theorem~\ref{thm.main_thm_noteBr} for analytic symbols, we need to introduce the notion of implementability. A symplectomorphism $\upphi : \hi \to \hi$ is said to be implementable if there exists a unitary operator $U$ on $\focks$, called a Bogoliubov implementer of $\upphi$, such that  
\begin{equation}\label{def_symplectomorphisme_implémentable_noteBr}
    U^{\dagger}\, \weylop[\eps]{v}\, U = \weylop[\eps]{\upphi v} \text{ for all } v \in \hi\,,
\end{equation}
where $\weylop[\eps]{v} = \e^{\frac{\rmi}{\sqrt{2}} \left(\anni[\eps]{v} + \crea[\eps]{v}\right)}$ denotes the Weyl operator associated with $v$ (here, $\anni[\eps]{v}$ and $\crea[\eps]{v}$ are the usual semi-classical annihilation and creation operators, see for instance \cite{AmNi08}). We state below a characterization of implementable symplectomorphisms due to D.~Shale. Recall that the adjoint of a $\C$-antilinear operator $A$ is the unique $\C$-antilinear operator $A^\dagger$ satisfying $\braket{z_1}{A\, z_2} = \braket{z_2}{A^\dagger\, z_1}$ for all $z_1,z_2 \in \hi$.
\begin{theorem}[\,\cite{Sh62}]\label{thm.Shale_implementable_symplecto_noteBr}
Let $\upphi$ be a bounded symplectomorphism on $\hi$, and let us write $\upphi$ as the sum of its $\C$-linear and $\C$-antilinear parts, that is, 
\begin{equation}\label{decomp_symplecto_linear_antilinear_noteBr}
    \upphi = L + A\,, \text{ where } L = \frac{1}{2}\, (\upphi - \rmi\, \upphi\, \rmi) \text{ and } A = \displaystyle\frac{1}{2}\, (\upphi + \rmi\, \upphi\, \rmi)\,.
\end{equation}
Then, the symplectomorphism $\upphi$ is implementable if and only if $A$ is Hilbert--Schmidt, in the sense that the $\C$-linear operator $A^\dagger\, A$ is trace class.
\end{theorem}
By Theorem~\ref{thm.Shale_implementable_symplecto_noteBr}, any bounded implementable symplectomorphism $\upphi$ admits a unique decomposition 
\begin{equation}\label{decomposition_symplectomorphisme_noteBr}
    \upphi = L + A\,, \text{ where } \left\{\begin{array}{l}
        \text{$L$ is $\C$-linear,}\\
        \text{$A$ is $\C$-antilinear, Hilbert--Schmidt.}
    \end{array}\right.
\end{equation}
On the set of implementable symplectomorphisms, denoted by $\implsym$, we can define the finite quantity 
\begin{equation}
    \norm{\upphi}_{\implsym} = \norm{L}_{\mathcal{L}(\hi)} + \norm{A}_{\mathcal{B}_2(\hi)}\,.
\end{equation}
Here, $\mathcal{B}_2(\hi)$ denotes Hilbert--Schmidt operators,
linear or antilinear, and $\norm{A}_{\mathcal{B}_2(\hi)}$ is the Hilbert--Schmidt norm of $A$, defined as the square root of the trace norm of $A^\dagger\, A$.

\medskip

Geometrically speaking, the antilinear part of a symplectomorphism $\upphi$ measures how much $\upphi$ deforms the complex structure on the Hilbert space $\hi$. For more details about $\R$-linear symplectic transformations, we refer the reader to \cite[Appendix~A]{Br12}.

\subsection{\texorpdfstring{Operator $\Lambda_t$ for analytic symbols}{Operator Lambda\_t for analytic symbols}}\label{subsect.op_Lambda_t_noteBr}

Let us now come back to the setting described in Section \ref{sect.main_result_noteBr} and prove that the operator $\Lambda_t$ given by \eqref{op_Lambda_t_noteBr} is well-defined pointwise on the class $\FVclass$ for all $t \in \R$. 

\medskip

\noindent\textbf{\large Step 1.} First, we provide precise definitions of the solutions to \eqref{classical_evolution_noteBr} and \eqref{quantum_evolution_noteBr}, following \cite[Sections 4 and 5]{Br12}. Recalling that $Q_t(z) = \braket{z}{\mathfrak{a}_t\, z} + \Im\braket{\mathfrak{b}_t}{\tensn{z}{2}}$ is the quadratic symbol \eqref{def_quadratic_symbol_noteBr}, we start by passing to the interaction representation in order to reduce to the case $\mathfrak{a}_t = 0$. Let us note that the classical dynamics generated by the free part $\braket{z}{\mathfrak{a}_t\, z}$ is $(u_\mathfrak{a}(t))_{t \in \R}$, the family of unitary operators satisfying \eqref{equation_alphat_noteBr}, and the corresponding quantum propagator is $\liftop{u_\mathfrak{a}(t)}$. 
Setting 
\begin{equation}\label{def_quadratic_symbol_interaction_RPZ_noteBr}
    \tilde{Q}_t(z) = \Im\braket{\tilde{\mathfrak{b}}_t}{\tensn{z}{2}}\,,
    \,\text{ where } \tilde{\mathfrak{b}}_t = \liftop{u_\mathfrak{a}(t)}^{-1}\mathfrak{b}_t \in C^0(\R,\tensymn{\hi}{2})\,,
\end{equation}
the operator $\gatbar{}{\tilde{Q}_t} = \rmi\, \left(\1 \otimes \bra{\bullet}\right)\, \tilde{\mathfrak{b}}_t$ is a $\C$-antilinear Hilbert--Schmidt operator on $\hi$ satisfying \[\norm{\gatbar{}{\tilde{Q}_t}-\gatbar{}{\tilde{Q}_s}}_{\mathcal{B}_2(\hi)} = \norm{\tilde{\mathfrak{b}}_t-\tilde{\mathfrak{b}}_s}_{\tensn{\hi}{2}}\,.\] Hence, $t \mapsto \gatbar{}{\tilde{Q}_t}$ is continuous in $\mathcal{B}_2(\hi)$ by continuity of $\mathfrak{b}_t$ and strong continuity of $u_\mathfrak{a}(t)$. This enables us to prove that the classical flow $\tilde{\upphi}_t$ of the Cauchy problem
\begin{equation}\label{classical_evolution_interaction_RPZ_noteBr}
	\left\{\begin{array}{l}
		\rmi\, \dot{\tilde{z}_t} = \gatbar{}{\tilde{Q}_t}(\tilde{z}_t)\\
		\eval{\tilde{z}_t}_{t=0} = z_0 \in \hi
		\end{array}\right.
\end{equation}
is well-defined for all $t \in \R$ by virtue of the Cauchy--Lipschitz theorem. Furthermore, by differentiating the expression \[t \mapsto \symform{\tilde{\upphi}_t\, z_1}{\tilde{\upphi}_t\, z_2}\] where $\symform{z_1}{z_2} = \Im\braket{z_1}{z_2}$ is the symplectic form of Subsection~\ref{subsect.evolution_analytic_symbols_noteBr}, we see that $\tilde{\upphi}_t$ is a symplectomorphism.

\medskip

We now turn to the quantum evolution. For all $k \in \N^*$, the problem \begin{equation}\label{quantum_evolution_interaction_RPZ_noteBr}
		\left\{\begin{array}{l}
			\rmi \eps\, \partial_t \tilde{\psi}_t = \Wick[\eps]{\tilde{Q}_t}\,\tilde{ \psi}_t\\
			\eval{\tilde{\psi}_t}_{t=0} = \psi_0
		\end{array}\right.
\end{equation} 
can be solved in the Hilbert rigging $D\left(\nbop^{k/2}\right) \subset \focks \subset D\left(\nbop^{k/2}\right)^*$, which provides a unique unitary propagator $\left(\tilde{U}_t\right)_{t \in \R}$: we refer to Appendix~\ref{app.proof_crucial_estimate_noteBr} for further discussion. Finally, the solutions of the classical dynamics \eqref{classical_evolution_noteBr} and the quantum evolution \eqref{quantum_evolution_noteBr} are respectively defined as 
\begin{equation}\label{dynamics_without_alphat_noteBr}
    \upphi_t = u_\mathfrak{a}(t)\, \tilde{\upphi}_t \text{ and } 
    U_t = \liftop{u_\mathfrak{a}(t)}\, \tilde{U}_t\,.
\end{equation}

\medskip

\noindent\textbf{\large Step 2.} Let us prove that the symplectomorphism $\tilde{\upphi}_t$ is implementable. Splitting $\tilde{\upphi}_t$ into its $\C$-linear and $\C$-antilinear parts as in Theorem~\ref{thm.Shale_implementable_symplecto_noteBr}, $\tilde{\upphi}_t = \tilde{L}_t + \tilde{A}_t$, and using the $\C$-antilinearity of $\gatbar{}{\tilde{Q}_t}$, the Cauchy problem \eqref{classical_evolution_interaction_RPZ_noteBr} is equivalent to the two systems
\[\left\{\begin{array}{l}
		\rmi\, \partial_t\, \tilde{L}_t = \gatbar{}{\tilde{Q}_t}\, \tilde{A}_t\\
		{}\\
		\eval{\tilde{L}_t}_{t=0} = \1
		\end{array}\right. \text{ and } \left\{\begin{array}{l}
		\rmi\, \partial_t\, \tilde{A}_t = \gatbar{}{\tilde{Q}_t}\, \tilde{L}_t\\
		{}\\
		\eval{\tilde{A}_t}_{t=0} = 0\,.
		\end{array}\right.\]
Writing the Duhamel formula for $\tilde{L}_t$ and $\tilde{A}_t$ and using $\norm{\gatbar{}{\tilde{Q}_t}}_{\mathcal{B}_2(\hi)} = \norm{\tilde{\mathfrak{b}}_t}_{\tensn{\hi}{2}}$, we obtain the norm estimates
\[\norm{\tilde{A}_t}_{\mathcal{B}_2(\hi)} \leq \int_{I_t} \norm{\tilde{\mathfrak{b}}_\tau}_{\tensn{\hi}{2}} \norm{\tilde{L}_\tau}_{\mathcal{L}(\hi)}\, \dd{\tau} < + \infty\]
and 
\[\norm{\tilde{L}_t}_{\mathcal{L}(\hi)} \leq 1 + \int_{I_t} \norm{\tilde{\mathfrak{b}}_\tau}_{\tensn{\hi}{2}}\, \norm{\tilde{A}_\tau}_{\mathcal{B}_2(\hi)}\, \dd{\tau}\,,\]
where $I_t = [\min(0,t), \max(0,t)]$. In particular, the antilinear part of $\tilde{\upphi}_t$ is Hilbert--Schmidt, so $\tilde{\upphi}_t$ is implementable by virtue of Theorem~\ref{thm.Shale_implementable_symplecto_noteBr}. Furthermore, adding the previous estimates and using Grönwall's lemma, we obtain 
\begin{equation}
    \norm{\tilde{\upphi}_t}_{\implsym} \leq \exp \int_{I_t} \norm{\tilde{\mathfrak{b}}_\tau}_{\tensn{\hi}{2}}\, \dd{\tau} = \exp \abs{\int_0^t \norm{\tilde{\mathfrak{b}}_\tau}_{\tensn{\hi}{2}}\, \dd{\tau}}\,.
\end{equation}
Consequently, since $u_\mathfrak{a}(t)$ is unitary and since $\norm{\mathfrak{b}_t}_{\tensn{\hi}{2}} = \norm{\tilde{\mathfrak{b}}_t}_{\tensn{\hi}{2}}$, the classical flow ${\upphi_t = u_\mathfrak{a}(t)\, \tilde{\upphi}_t}$ is also implementable and satisfies the norm estimate 
\begin{equation}\label{Gronwall_estimate_classical_flow_noteBr}
    \norm{\upphi_t}_{\implsym} \leq \exp \abs{\int_0^t \norm{\mathfrak{b}_\tau}_{\tensn{\hi}{2}}\, \dd{\tau}}\,.
\end{equation}

\bigskip

\noindent\textbf{\large Step 3.} The decomposition \eqref{decomposition_symplectomorphisme_noteBr} enables us to define the operator $\Lambda_t$ pointwise on the space of polynomial symbols $\wickpol$ (see Subsection \ref{subsect.math_background_noteBr}) by the expression \eqref{op_Lambda_t_noteBr}, rewritten below for the reader's convenience:
\begin{equation*}
    \Lambda_t b(z) = -2\, \Tr{A_t^{\dagger} A_t\, \gatgatbar{}{}{b}(z)} + \braket{v_t}{\gatbar{2}{b}(z)}_{\tensn{\hi}{2}} + \braket{\gat{2}{b}(z)}{v_t}_{(\tensn{\hi}{2})^*,\tensn{\hi}{2}}\,.
\end{equation*}
We recall that the Gâteaux differentials $\gat{2}{b}(z)$, $\gatbar{2}{b}(z)$ and $\gatgatbar{}{}{b}(z)$ are given by \eqref{def_gatgatbar_noteBr}, and $v_t \in \tensn{\hi}{2}$ is the unique vector satisfying $\braket{z_1 \ot z_2}{v_t} = \braket{z_1}{L_t^{\dagger}\, A_t\, z_2}$ for all $z_1,z_2 \in \hi$, following \cite[Theorem~3.2]{Br12}.

\medskip

Let $V \in \focks$ be fixed. Differentiating the series \eqref{symb_Fvclass_noteBr} term by term provides the second-order Gâteaux differentials of $F_V$ in the sense of \eqref{dév_limité_dérivées_Fréchet_polynôme_Wick_noteBr}: see \eqref{computation_gatbar_Fv_noteBr}, \eqref{computation_gat_Fv_noteBr} and \eqref{computation_gatgatbar_Fv_noteBr} below for the explicit expressions, given by convergent series by virtue of the Cauchy--Schwarz inequality. Finally, the expression $\Lambda_t F_V(z)$ is well-defined for all $z \in \hi$.
	 
\subsection{\texorpdfstring{Exponential operator associated with $\Lambda_t$}{Exponential operator associated with Lambda\_t}}\label{subsect.exp_op_Lambda_t_noteBr}

Let us now prove that the exponential series $\e^{\frac{\eps}{2} \Lambda_t} F_V$ defines an element of $\smallFVclass$ for all ${F_V \in \smallFVclass}$ and $0 < \eps \leq \eps(t)$ sufficiently small (see \eqref{def_semiclassical_param_noteBr} below).

\medskip

First, for $F_V \in \smallFVclass$, let us explicitly compute $\Lambda_t F_V(z)$ for all $z \in \hi$. We start by computing the Gâteaux differentials of $F_V$:
	    \begin{equation}\label{computation_gatbar_Fv_noteBr}
	        \gatbar{2}{F_V}(z) = \sum_{j=2}^\infty \frac{j\,(j-1)}{\sqrt{j!}}\,\Big((\tensn{\1}{2}) \ot \bra{\tensn{(z+\conjug z)}{j-2}}\Big)\,V^{(j)} \in \tensymn{\hi}{2}\,,
	    \end{equation}
	    \begin{equation}\label{computation_gat_Fv_noteBr}
	        \gat{2}{F_V}(z) = \sum_{j=2}^\infty \frac{j\,(j-1)}{\sqrt{j!}}\,\braket{\liftop{\conjug}\bullet}{\Big((\tensn{\1}{2}) \ot \bra{\tensn{(z+\conjug z)}{j-2}}\Big)\,V^{(j)}} \in (\tensymn{\hi}{2})^*\,,
	    \end{equation}
	    and 
	    \begin{equation}\label{computation_gatgatbar_Fv_noteBr}
	        \gatgatbar{}{}{F_V}(z) = \sum_{j=2}^\infty \frac{j\,(j-1)}{\sqrt{j!}}\,\Big(\1 \ot \bra{\conjug\, \bullet}\Big)\,\Big((\tensn{\1}{2}) \ot \bra{\tensn{(z+\conjug z)}{j-2}}\Big)\,V^{(j)} \in \mathcal{L}(\hi)\,.
	    \end{equation}
Let us denote by $(e_q)_{q \in \N}$ an orthonormal basis of $\hi$ such that $\conjug\, e_q = e_q$ for all $q \in \N$: it suffices to pick an orthonormal basis of the real Hilbert space $\hi_\conjug = \{z \in \hi\,,\, \conjug z = z\}$, since we have the decomposition $\hi = \hi_\conjug \oplus \rmi\, \hi_\conjug$. Then, the expression \eqref{op_Lambda_t_noteBr} yields 
	    \begin{multline}\label{op_Lambda_t_Fv}
	        \Lambda_t F_V(z) = \sum_{j=2}^\infty \frac{j\,(j-1)}{\sqrt{j!}} \times\\ \braket{-2\,\sum_{q \in \N}(A_t^{\dagger} A_t\, e_q \ot e_q) + \liftop{\conjug}v_t+v_t}{\Big((\tensn{\1}{2}) \ot \bra{\tensn{(z + \conjug z)}{j-2}}\Big)\, V^{(j)}}
	    \end{multline}
for all $F_V \in \smallFVclass$, where the vector 
\begin{equation}
    Y_t \defegal -2\,\sum_{q \in \N} (A_t^{\dagger} A_t\, e_q \ot e_q) + \liftop{\conjug}v_t+v_t \in \tensn{\hi}{2}\,
\end{equation}
	    is well-defined since $A_t$ is Hilbert--Schmidt (see \eqref{decomposition_symplectomorphisme_noteBr}). Using the fact that \[\displaystyle\frac{j\, (j-1)}{\sqrt{j!}} = \displaystyle\frac{\sqrt{j\, (j-1)}}{\sqrt{(j-2)!}}\] and re-indexing the sum in \eqref{op_Lambda_t_Fv}, we obtain that the symbol $\Lambda_t F_V$ belongs to the Hilbert space $\FVclass$ whenever $V \in \focks$ satisfies Assumption~\ref{hyp.potentiel_petit_noteBr}, with associated potential $V_t$ defined by
	    \begin{equation}\label{pot_LamTFv_noteBr}
	        V_t^{(j)} = \sqrt{(j+1)\,(j+2)}\, \left((\tensn{\1}{j}) \ot \bra{Y_t}\right)\, V^{(j+2)},\quad \forall\,j \in \N\,.
	    \end{equation}
	    Let us note that the vector $v_t \in \tensn{\hi}{2}$ appearing in formula \eqref{op_Lambda_t_noteBr} satisfies \[\norm{v_t}_{\tensn{\hi}{2}} = \norm{L_t^{\dagger}\,A_t}_{\mathcal{B}_2(\hi)} \leq \norm{L_t^{\dagger}}_{\mathcal{L}(\hi)}\, \norm{A_t}_{\mathcal{B}_2(\hi)} \leq \norm{\upphi_t}_{\implsym}^2\,,\] where $\upphi_t = L_t + A_t$ is the decomposition of the classical flow provided by \eqref{decomposition_symplectomorphisme_noteBr}. Combined with the estimate $\norm{A_t^\dagger\, A_t}_{\mathcal{B}_2(\hi)} \leq \norm{A_t}_{\mathcal{B}_2(\hi)}^2 \leq \norm{\upphi_t}_{\implsym}^2$ and the fact that $\liftop{\conjug}$ is isometric, this leads to the estimate \[\norm{Y_t}_{\tensn{\hi}{2}} \leq 4\, \norm{\upphi_t}_{\implsym}^2\,.\] Then, coming back to \eqref{pot_LamTFv_noteBr}, we obtain 
	    \begin{equation}\label{estimee_V1_noteBr} \norm{V_t^{(j)}}_{\tensn{\hi}{j}}^2 \leq \left(4 \norm{\upphi_t}_{\implsym}^2\right)^2 \, (j+1)\,(j+2)\, \norm{V^{(j+2)}}_{\tensn{\hi}{j+2}}^2\,.
	    \end{equation}
	    Using $\e^{2\gamma\lambda^j}(j+1)(j+2)\leq C\,\e^{2\gamma\lambda^{j+2}}$ for some $C = C(\gamma,\lambda)$ independent of $j$, where $\gamma > 0$ and $\lambda > 1$ are the constants of Assumption~\ref{hyp.potentiel_petit_noteBr}, we obtain that $\Lambda_t F_V \in \smallFVclass$. Consequently, the operator $\Lambda_t$ preserves the class $\smallFVclass$.

\medskip

For all $k \in \N^*$, by iterating \eqref{pot_LamTFv_noteBr} $k$ times, we see that the potential $V_{t,k}$ associated with $\Lambda_t^k F_V$ is given by
\begin{multline}
    V_{t,k}^{(j)} = \sqrt{(j+1)\cdots(j+2k)}\, \left((\tensn{\1}{j}) \ot \bra{Y_t}\right)\cdots\left((\tensn{\1}{j+2(k-1)}) \ot \bra{Y_t}\right)\, V^{(j+2k)}\,,\\\quad \forall\,j \in \N\,.
\end{multline}
The estimate $(j+1)\cdots(j+2k) = (2k)!\, \displaystyle\binom{j+2k}{j} \leq (2k)!\, 2^{j+2k}$ provides
\begin{equation}\label{estimee_Vk_noteBr}
    \norm{V_{t,k}^{(j)}}_{\tensn{\hi}{j}}^2 \leq \left(4 \norm{\upphi_t}_{\implsym}^2\right)^{2k}\, (2k)!\, 2^{j+2k}\, \norm{V^{(j+2k)}}_{\tensn{\hi}{j+2k}}^2\,,
\end{equation}
which yields, summing over $j$, 
\begin{equation}\label{estimee_Lambda_T_k_noteBr}
	   \norm{\Lambda_t^k F_V}_{\FVclass} \leq \left(4 \norm{\upphi_t}_{\implsym}^2\right)^k \, (2k)!^{1/2}\, \norm{\liftop{\sqrt{2}}V}_{\focks}\,,\quad \forall\ k \in \N\,.
\end{equation}
	    Summing these estimates over $k$ and using
	    \begin{equation}\label{formule_Stirling_noteBr}
	        \displaystyle\frac{(2k)!^{1/2}}{k!} = \binom{2k}{k}^{1/2} \leq 2^k\,,
	    \end{equation}
	    we have   
	 \begin{align*}
	     \sum_{k=0}^\infty \left(\frac{\eps}{2}\right)^k\, \frac{1}{k!}\, \norm{\Lambda_t^k F_V}_{\FVclass} \leq &\sum_{k=0}^\infty \left(4\,\eps\,\norm{\upphi_t}_{\implsym}^2\right)^k \norm{\liftop{\sqrt{2}}V}_{\focks}\\
	     \leq &2\, \norm{\liftop{\sqrt{2}}V}_{\focks} \quad \text{if $4\,\eps\, \norm{\upphi_t}_{\implsym}^2 \leq \displaystyle\frac{1}{2}$}\,.
	 \end{align*}
	 Hence, for all $F_V \in \smallFVclass$, the series \[\displaystyle\sum \frac{\eps^k}{2^k\, k!}\, \Lambda_t^k F_V\] converges in the Hilbert space $\FVclass$ for $0 < \eps \leq \eps(t)$, where 
	 \begin{equation}\label{def_semiclassical_param_noteBr}
	     \eps(t) \egaldef \displaystyle\frac{1}{8\, \norm{\upphi_t}_{\implsym}^2}\,.
	 \end{equation}
	 Its sum, denoted by $\e^{\frac{\eps}{2} \Lambda_t} F_V$, satisfies the estimate 
	\begin{equation}\label{estimee_expo_Fv_noteBr}
	    \norm{\e^{\frac{\eps}{2} \Lambda_t} F_V}_{\FVclass} \leq 2\, \norm{\liftop{\sqrt{2}} V}_{\focks}\,.
	\end{equation}
	
	\medskip
	
	Let us now prove that the symbol $\e^{\frac{\eps}{2} \Lambda_t} F_V$ belongs to $\smallFVclass$. Writing $\e^{\frac{\eps}{2} \Lambda_t} F_V$ in the form $F_{\tilde{V}_{t,\eps}}$ with $\tilde{V}_{t,\eps} \in \focks$, we deduce from \eqref{estimee_Vk_noteBr} and \eqref{formule_Stirling_noteBr} that
	\begin{align*}
    \norm{\Tilde{V}_{t,\eps}^{(j)}}_{\tensn{\hi}{j}}\leq &\sum_{k = 0}^\infty \frac{\eps^k}{2^k\, k!}\, \norm{V_{t,k}^{(j)}}_{\tensn{\hi}{j}}\\
    \leq &\sum_{k = 0}^\infty \frac{\eps^k}{2^k\, k!}\, \left(4\, \norm{\upphi_t}_{\implsym}^2\right)^{k}\, (2k)!^{1/2}\,\left(\sqrt{2}\right)^{j+2k}\,\norm{V^{(j+2k)}}_{\tensn{\hi}{j+2k}}\\
    = &\sum_{k = 0}^\infty \left(2\, \eps\,\norm{\upphi_t}_{\implsym}^2\right)^k\,\frac{(2k)!^{1/2}}{k!}\,\left(\sqrt{2}\right)^{j+2k}\,\norm{V^{(j+2k)}}_{\tensn{\hi}{j+2k}}\\
    \leq\, &C \sum_{k=0}^\infty \left(4\,\eps\, \norm{\upphi_t}_{\implsym}^2\right)^k\, (\sqrt{2})^{j+2k}\, \norm{V^{(j+2k)}}_{\tensn{\hi}{j+2k}}\,,
\end{align*}
	where $C > 0$ is an $\eps$-independent constant which may vary from one estimate to another in what follows. This yields, by the Cauchy--Schwarz inequality, 
	\begin{equation}
    \norm{\Tilde{V}_{t,\eps}^{(j)}}_{\tensn{\hi}{j}}^2 \leq C\, \sum_{k=0}^\infty 2^{j+2k} \norm{V^{(j+2k)}}_{\tensn{\hi}{j+2k}}^2\,
	\end{equation}
	for $0 < \eps \leq \eps(t)$. Recall that $V \in D\left(\e^{\gamma \liftop{\lambda}}\right)$ for some $\gamma > 0$ and $\lambda > 1$. Using the estimate above, we obtain, for all $1 < \lambda' < \lambda$, 
	\begin{align*}
	    \norm{\e^{\gamma \liftop{\lambda'}} \tilde{V}_{t,\eps}}_{\focks}^2 = &\sum_{j=0}^\infty \e^{2 \gamma (\lambda')^j} \norm{\tilde{V}_{t,\eps}^{(j)}}_{\tensn{\hi}{j}}^2\\
	    \leq &\,C\, \sum_{j = 0}^\infty \sum_{k = 0}^\infty \e^{2 \gamma (\lambda')^j}\, 2^{j+2k}\, \norm{V^{(j+2k)}}_{\tensn{\hi}{j+2k}}^2\\
	    = &\,C\, \sum_{k = 0}^\infty \sum_{\ell = 2k}^\infty \e^{2 \gamma (\lambda')^{\ell-2k}}\, 2^\ell\, \norm{V^{(\ell)}}_{\tensn{\hi}{\ell}}^2\,,
	\end{align*}
	where we have set $\ell=j+2k$. By Tonelli's theorem and the fact that the factor $\displaystyle\sum_{k=0}^{\ent{\ell/2}} 2^\ell\, \e^{2 \gamma (\lambda')^{\ell-2k}}$ is bounded by $C\,\e^{2 \gamma \lambda^\ell}$, we obtain that $\tilde{V}_{t,\eps}$ satisfies Assumption~\ref{hyp.potentiel_petit_noteBr} with constants $\gamma$ and $\lambda'$. Hence, the symbol $\e^{\frac{\eps}{2} \Lambda_t} F_V = F_{\tilde{V}_{t,\eps}}$ belongs to $\smallFVclass$.

\subsection{Composition with the classical flow}\label{subsect.compo_classical_flow_noteBr}

At this point, we know that each operator $\e^{\frac{\eps}{2} \Lambda_t}$ preserves the class $\smallFVclass$ for all ${0 < \eps \leq \eps(t)}$ (see \eqref{def_semiclassical_param_noteBr}), and that $\norm{\e^{\frac{\eps}{2} \Lambda_t}F_V}_{\FVclass}$ is dominated by $2\, \norm{\liftop{\sqrt{2}}V}_{\focks}$ (see \eqref{estimee_expo_Fv_noteBr}). In formula \eqref{generalisation_formule_Breteaux_conj_Wick_par_propag_quad}, the exponential operator $\e^{\frac{\eps}{2} \Lambda_t}$ is applied to $F_V \circ \upphi_t$, which does not belong to $\FVclass$ in general. However, we note that Subsection~\ref{subsect.implementability_noteBr} does not involve the complex conjugation $\conjug$, and that the arguments used in Subsections \ref{subsect.op_Lambda_t_noteBr} and \ref{subsect.exp_op_Lambda_t_noteBr} can be used with $\conjug$ replaced by any other complex conjugation without affecting the conclusions. We state below a crucial result, whose proof is provided in Appendix~\ref{app.time_dep_conjugations_noteBr}.
\begin{prop}\label{prop.time_dep_conjugations_noteBr}
     For all $t \in \R$, let $X_t = L_t^\dagger + A_t^\dagger\, \conjug$. Then, $X_t$ is invertible, and the map ${\conjug_t = X_t\, \conjug\, X_t^{-1}}$ defines a complex conjugation on $\hi$. Furthermore, we have the identity \[(\1 + \conjug)\, \upphi_t = X_t^{\dagger}\, (\1 + \conjug_t)\,.\]
\end{prop}

Let us note that, if $V \in \focks$ satisfies Assumption~\ref{hyp.potentiel_petit_noteBr} for some constants $\gamma > 0$ and $\lambda > 1$, we have, for all $0 < \gamma' < \gamma$,
\begin{equation}
    \norm{\e^{\gamma' \liftop{\lambda}}\, \liftop{X_t}\, V}_{\focks}^2 \leq \left(\underset{j \in \N}{\sup}\, \e^{-2(\gamma-\gamma')\lambda^j}\, \norm{X_t}_{\mathcal{L}(\hi)}^{2j}\right)\, \norm{\e^{\gamma \liftop{\lambda}} V}_{\focks}^2 < +\infty\,
\end{equation}
where the sup is finite since $\norm{X_t}_{\mathcal{L}(\hi)} \leq \norm{L_t}_{\mathcal{L}(\hi)} + \norm{A_t}_{\mathcal{B}_2(\hi)} \leq \norm{\upphi_t}_{\implsym}$. Hence, $\liftop{X_t}\, V$ also satisfies Assumption~\ref{hyp.potentiel_petit_noteBr} with constants $\gamma' > 0$ and $\lambda > 1$.

Now, let $\FVclass[t]$ denote the class of symbols of the form \eqref{symb_Fvclass_noteBr} with $\conjug$ replaced by $\conjug_t$, endowed with the inner product 
\begin{equation*}
    \braket{F_{V_1}^{\conjug_t}}{F_{V_2}^{\conjug_t}}_{\FVclass[t]} = \braket{V_1}{V_2}_{\focks}\,.
\end{equation*}
According to Proposition~\ref{prop.time_dep_conjugations_noteBr}, for all $F_V \in \smallFVclass$, we have $F_V \circ \upphi_t \in \FVclass[t]$, and the associated potential is $\liftop{X_t}\, V$. Reproducing the arguments of the previous subsections, we obtain that the symbol 
\begin{equation}\label{exp_lambda_t_Fv_phi_t_noteBr}
	   \e^{\frac{\eps}{2} \Lambda_t} (F_V \circ \upphi_t)
\end{equation}
belongs to $\FVclass[t]$ whenever $F_V \in \smallFVclass$ and $0 < \eps \leq \eps(t)$, with the estimate 
\begin{align}
    \norm{\e^{\frac{\eps}{2} \Lambda_t} (F_V \circ \upphi_t)}_{\FVclass[t]} \leq &2\, \norm{\liftop{\sqrt{2}X_t}V}_{\focks}\label{estimee_expo_Fv_circ_flot_noteBr}\\
    \leq &2\, \left(\underset{j \in \N}{\sup}\, \e^{-2\gamma\lambda^j}\, 2^j\, \norm{X_t}_{\mathcal{L}(\hi)}^{2j}\right)^{1/2}\, \norm{\e^{\gamma \liftop{\lambda}}\, V}_{\focks}\,.\nonumber
\end{align}
We note that the condition $4\, \eps\, \norm{\upphi_t}_{\implsym}^2 \leq \displaystyle\frac{1}{2}$ remains unchanged.

\subsection{End of the proof}\label{subsect.end_proof_noteBr}

We can now conclude the proof of Theorem~\ref{thm.main_thm_noteBr} in three steps:
\begin{enumerate}
    \item We start by considering truncations of the potential of $F_V \in \smallFVclass$, to which we apply \cite[Theorem~3.2]{Br12}. Thanks to this strategy, the underlying geometric setting for analytic symbols is exactly the one developed in \cite{Br12} for polynomial symbols.
    \item Then, we pass to the limit and prove a version of \eqref{generalisation_formule_Breteaux_conj_Wick_par_propag_quad} in the strong sense on the finite-particle space $\focksalg$. The mechanism at stake combines hypercontractivity estimates with Proposition~\ref{prop.inegalite_cruciale_reste_propag_CS_QFT_noteBr}, adapted from \cite[Lemma~4.9]{AmZe14} and proved in Appendix~\ref{app.proof_crucial_estimate_noteBr}, where the super-exponential decay assumption \ref{hyp.potentiel_petit_noteBr} plays a crucial role.
    \item Finally, we conclude the proof of Theorem~\ref{thm.main_thm_noteBr} by density arguments.
\end{enumerate}

\bigskip

\noindent\textbf{\large Step 1.} We work on the compact interval $t \in [-t_0,t_0]$ with $t_0 > 0$ fixed, and we define 
\[
M(t_0) = \displaystyle\underset{\abs{t} \leq t_0}{\sup}\, \norm{\upphi_t}_{\implsym}\,.
\]
This quantity is finite and dominated by $\exp\left(\displaystyle\int_{-t_0}^{t_0} \norm{\mathfrak{b}_\tau}_{\tensn{\hi}{2}}\,\dd{\tau}\right)$, see \eqref{Gronwall_estimate_classical_flow_noteBr}.

Let $V \in \focks$ be fixed, fulfilling Assumption~\ref{hyp.potentiel_petit_noteBr} with constants $\gamma > 0$ and $\lambda > 1$. For all $m \in \N$, let us write $V = V_m + R_m$ where $V_m = \displaystyle\bigoplus_{j=0}^m V^{(j)}$ and $R_m = \displaystyle\bigoplus_{j=m+1}^\infty V^{(j)}$. Since $F_{V_m}$ is a polynomial symbol of order at most $m$, the symbol \eqref{exp_lambda_t_Fv_phi_t_noteBr} associated with $F_{V_m}$ is simply given by the finite sum \begin{equation}\label{formule_symbole_distordu_par_propagation_quad}
	    \e^{\frac{\eps}{2} \Lambda_t} (F_{V_m} \circ \upphi_t) = \sum_{k=0}^{\ent{m/2}} \frac{\eps^k}{2^k k!}\, \Lambda_t^k (F_{V_m} \circ \upphi_t)\,.
	\end{equation}
	 Furthermore, coming back to the quadratic symbol \eqref{def_quadratic_symbol_noteBr}, we recall that the operators $\mathfrak{a}_t$ satisfy the assumptions of \cite[Theorem~X.70]{ReSi75} (see Subsection~\ref{subsect.evolution_analytic_symbols_noteBr}) and that $\mathfrak{b}_t \in C^0(\R,\tensymn{\hi}{2})$. Then, the hypotheses of \cite[Theorem~3.2]{Br12} are fulfilled. We also note that the procedure \cite[Subsection~3.2]{AmZe14} defining the Wick quantization of analytic symbols in $\FVclass$ also applies to the class $\FVclass[t]$ defined in Subsection~\ref{subsect.compo_classical_flow_noteBr}, with $\conjug$ replaced by $\conjug_t = X_t\, \conjug\, X_t^{-1}$ (see Proposition~\ref{prop.time_dep_conjugations_noteBr}): in particular, the Wick quantization on $\FVclass$ agrees with polynomial Wick quantization on the space ${\wickpol \cap \FVclass}$, and the same holds for the class $\FVclass[t]$. Consequently, we can apply \cite[Theorem~3.2]{Br12} to the polynomial symbol $F_{V_m}$, which provides 
	 \begin{equation}\label{formule_Breteaux_conj_Wick_par_propag_quad}
			U_t^{\dagger}\, \Wick[\eps]{F_{V_m}}\, U_t = \Wick[\eps]{\e^{\frac{\eps}{2} \Lambda_t} (F_{V_m} \circ \upphi_t)}
		\end{equation}
	in the space of bounded operators $\mathcal{L}\left(D\Big(\nbop^{m/2}\Big),\focks\right)$. Here, $\nbop$ denotes the number operator defined by \eqref{nbop1_noteBr}.

\bigskip

\noindent\textbf{\large Step 2.} Let us take $\eps_*(t_0)\egaldef\underset{\abs{t}\leq t_0}{\inf}\,\eps(t) > 0$. Note that we have $\eps_*(t_0) \leq 1/8$ by virtue of \eqref{def_semiclassical_param_noteBr}, since ${\norm{\upphi_t}_{\implsym} \geq \norm{L_t}_{\mathcal{L}(\hi)} \geq 1}$ (see the proof of Lemma~\ref{lemma.time_dep_conjugations_noteBr}). For $0 < \eps \leq \eps_*(t_0)$ fixed, we wish to let $m \to + \infty$ in \eqref{formule_Breteaux_conj_Wick_par_propag_quad}, in order to obtain formula \eqref{generalisation_formule_Breteaux_conj_Wick_par_propag_quad} in the strong sense on the finite-particle space $\focksalg$ defined in \eqref{focksalg_noteBr}.

\medskip

We first focus on the right-hand side of \eqref{formule_Breteaux_conj_Wick_par_propag_quad}. For all $V \in \focks$ and ${\psi \in D\Big(\liftop{\sqrt{3}}\Big)}$, \cite[Lemma~3.14]{AmZe14} yields 
\begin{equation}\label{hyperc_estimate_noteBr}
    \norm{\Wick[\eps]{F_V} \psi}_{\focks} \leq \norm{V}_{\focks} \norm{\liftop{\sqrt{3}} \psi}_{\focks}
\end{equation}
for all $0 < \eps \leq 1/3$. This result is a consequence of the well-known hypercontractivity estimates \cite[Theorem~I.17]{Si74} used in the $Q$-space representation of Fock space (see for instance \cite[Proposition~2.1.1]{GlJa85} and also \cite[Theorem~I.11]{Si74}). More precisely, the $Q$-space representation provides a unitary realization $\mathcal{R} : \focks \to L^2(Q,\dd{\mu})$ of the bosonic Fock space $\focks$ as an $L^2$-space over a Gaussian probability space $(Q,\dd{\mu})$, which offers a convenient framework in which
field operators act as multiplication
operators. In this representation, for all $V \in \focks$, the operator $\Wick[\eps]{F_V}$ is the multiplication operator by $\mathcal{R}(\liftop{\sqrt{\eps}}V)$, which is real-valued if $\liftop{\conjug}V=V$ (see \cite{AmZe14}).

\bigskip

Since \cite[Lemma~3.14]{AmZe14} holds for any complex conjugation on $\hi$ (the $Q$-space
representation may be built from $\conjug_t$), the hypercontractivity estimate \eqref{hyperc_estimate_noteBr} can be applied to the symbol $\e^{\frac{\eps}{2}\Lambda_t} (F_{R_m} \circ \upphi_t) \in \FVclass[t]$. Combined with the estimate \eqref{estimee_expo_Fv_circ_flot_noteBr}, this yields 
\begin{equation}\label{hyperc_estimate_reste_noteBr}
    \norm{\Wick[\eps]{\e^{\frac{\eps}{2}\Lambda_t} (F_{R_m} \circ \upphi_t)} \psi}_{\focks} \leq  C\, \norm{\liftop{\sqrt{2}\,M(t_0)} R_m}_{\focks} \norm{\liftop{\sqrt{3}} \psi}_{\focks}
\end{equation}
for all $\psi \in D\Big(\liftop{\sqrt{3}}\Big)$ and $0 < \eps \leq \min(\eps_*(t_0),1/3) = \eps_*(t_0)$ (recall that $\eps_*(t_0) \leq 1/8$). The right-hand side converges to $0$ as $m \to + \infty$ thanks to Assumption~\ref{hyp.potentiel_petit_noteBr}.

\medskip

Let us now turn our attention to the left-hand side of \eqref{formule_Breteaux_conj_Wick_par_propag_quad}. We have at hand the following crucial estimate, inspired by \cite[Lemma~4.9]{AmZe14}:
\begin{prop}\label{prop.inegalite_cruciale_reste_propag_CS_QFT_noteBr}
	 Let $(U_t)_{t \in \R}$ be the time-dependent unitary propagator of the evolution problem \eqref{quantum_evolution_noteBr}. Let $V \in \focks$ satisfy Assumption~\ref{hyp.potentiel_petit_noteBr}, that is, $V \in D\left(\e^{\gamma \liftop{\lambda}}\right)$ for some $\lambda > 1$ and $\gamma > 0$. Fix $1 < \lambda_0 < \lambda$ and $0 < \gamma_0 < 2\gamma/\lambda^2$. For all $\psi \in \focksalg$, $t \in \R$ and $0 < \eps \leq 1/3$, the state $U_t\, \psi$ belongs to $D\left(\Wick[\eps]{F_V}\right)$, and there exist constants $C,c_0 > 0$, depending only on $\gamma,\lambda,\gamma_0,\lambda_0$, such that we have 
	 \begin{multline*}
		\norm{\Wick[\eps]{F_V} \, U_t \, \psi}_{\focks} \leq C\, \norm{\e^{\gamma \liftop{\lambda}} \liftop{\sqrt{\eps}} V}_{\focks}\times\\
			\left(\norm{\Big(\mathrm{g}_t(\nbop+1)\Big)^{1/2}\, \psi}_{\focks}^2 + 2\, c_0 \, \mathrm{g}_t'(c_0) \abs{\int_0^t \norm{\mathfrak{b}_\tau}_{\tensn{\hi}{2}} \dd{\tau}} \norm{\psi}_{\focks}^2\right)^{1/2}\,.
		\end{multline*}
		Here, $\mathrm{g}_t$ is given by \begin{equation*}\label{def_serie_entiere_dependant_du_temps_noteBr}
			\mathrm{g}_t(x) = \sum_{k = 0}^\infty \e^{-\gamma_0 \, \lambda^k} \e^{2 \, k\, \lambda_0^k \abs{\int_0^t \norm{\mathfrak{b}_\tau}_{\tensn{\hi}{2}} \dd{\tau}}}\, x^k\,,
		\end{equation*}
		and $\mathrm{g}_t'(x) = \displaystyle\dv{x}\, \mathrm{g}_t(x)$. We recall that ${\mathfrak{b}_t \in C^0(\R,\tensymn{\hi}{2})}$ is the continuous family of vectors appearing in the definition of the quadratic symbol \eqref{def_quadratic_symbol_noteBr}.
	\end{prop}
	We refer to Appendix~\ref{app.proof_crucial_estimate_noteBr} for a detailed proof of Proposition~\ref{prop.inegalite_cruciale_reste_propag_CS_QFT_noteBr}. As stated in Remark~\ref{rem.on_main_thm_noteBr}, this proposition is where the super-exponential decay assumption \ref{hyp.potentiel_petit_noteBr} is useful, since the upper bounds on the norms $\norm{U_t}_{\mathcal{L}(D\left(\nbop^{k/2}\right))}$ obtained by the Cauchy--Schwarz and Grönwall estimates grow super-exponentially fast in $k$, see Proposition~\ref{prop.prop_4.5_AmZe_noteBr}.

We can apply Proposition~\ref{prop.inegalite_cruciale_reste_propag_CS_QFT_noteBr} to the potential $R_m = \displaystyle\bigoplus_{j=m+1}^\infty V^{(j)}$ instead of $V$, which satisfies Assumption~\ref{hyp.potentiel_petit_noteBr} with the same constants $\gamma > 0$, $\lambda > 1$. This yields an estimate of the form
\begin{multline}\label{inegalite_cruciale_reste_propag_CS_QFT_noteBr}
			\norm{U_t^{\dagger}\,\Wick[\eps]{F_{R_m}} \, U_t \, \psi}_{\focks} \leq C\, \eps^{\frac{m+1}{2}}\, \norm{\e^{\gamma \liftop{\lambda}}\, R_m}_{\focks}\times\\
			\left(\norm{\Big(\tilde{\mathrm{g}}_{t_0}(\nbop+1)\Big)^{1/2}\, \psi}_{\focks}^2 + 2\, c_0 \, \tilde{\mathrm{g}}_{t_0}{'}(c_0) \int_{-t_0}^{t_0} \norm{\mathfrak{b}_\tau}_{\tensn{\hi}{2}} \dd{\tau} \cdot \norm{\psi}_{\focks}^2\right)^{1/2}
		\end{multline}
		for all $t \in [-t_0,t_0]$, $0 < \eps \leq \min(\eps_*(t_0),1/3) = \eps_*(t_0)$ and $\psi \in \focksalg$, where 
		\begin{equation}
			\tilde{\mathrm{g}}_{t_0}(x) = \sum_{k = 0}^\infty \e^{-\gamma_0 \, \lambda^k} \e^{2 \, k\, \lambda_0^k \int_{-t_0}^{t_0} \norm{\mathfrak{b}_\tau}_{\tensn{\hi}{2}} \dd{\tau}}\, x^k\,,\quad 1 < \lambda_0 < \lambda\,,\quad 0 < \gamma_0 < \frac{2\gamma}{\lambda^2}\,.
		\end{equation}

\bigskip

Finally, using estimates \eqref{hyperc_estimate_reste_noteBr} and \eqref{inegalite_cruciale_reste_propag_CS_QFT_noteBr}, we can let $m \to +\infty$ in \eqref{formule_Breteaux_conj_Wick_par_propag_quad} and obtain 
\begin{equation}\label{generalisation_formule_Breteaux_conj_Wick_par_propag_quad_focksalg}
    U_t^{\dagger}\, \Wick[\eps]{F_V}\, U_t\, \psi = \Wick[\eps]{\e^{\frac{\eps}{2}\, \Lambda_t} (F_V \circ \upphi_t)}\, \psi
\end{equation}
for all $t \in [-t_0,t_0]$, $0 < \eps \leq \eps_*(t_0)$ and $\psi \in \focksalg$.

\bigskip

\noindent\textbf{\large Step 3.} To conclude the proof of Theorem~\ref{thm.main_thm_noteBr}, we remark that the domain $D\Big(\liftop{\sqrt{3}}\Big)$ is embedded in $D\Big(\left(\tilde{\mathrm{g}}_{t_0}(\nbop+1)\right)^{1/2}\Big)$. This follows from the identity
\begin{align*}
    \norm{\Big(\tilde{\mathrm{g}}_{t_0}(\nbop+1)\Big)^{1/2}\, \psi}_{\focks}^2 = \sum_{j=0}^\infty \tilde{\mathrm{g}}_{t_0}(j+1)\, 3^{-j}\, 3^{j} \norm{\psi^{(j)}}_{\tensn{\hi}{j}}^2
\end{align*}
combined with the fact that the quantities $(j+1)^k\, 3^{-j}$ are dominated uniformly in $j$ by \[\left(\displaystyle\frac{k}{\log 3}\right)^k\, \e^{\log 3-k}\,,\] and this bound is absorbed by the super-exponentially decaying
coefficients of $\tilde{\mathrm{g}}_{t_0}$. Thus, the estimates \eqref{hyperc_estimate_reste_noteBr} and \eqref{inegalite_cruciale_reste_propag_CS_QFT_noteBr} both extend to $\psi \in D\Big(\liftop{\sqrt{3}}\Big)$ by closedness of $\Wick[\eps]{F_V}$ and density, which provides \eqref{generalisation_formule_Breteaux_conj_Wick_par_propag_quad_focksalg} for all $\psi \in D\Big(\liftop{\sqrt{3}}\Big)$. Finally, we obtain that \eqref{generalisation_formule_Breteaux_conj_Wick_par_propag_quad} holds in the space $\mathcal{L}\left(D\Big(\liftop{\sqrt{3}}\Big), \focks\right)$, which concludes the proof of Theorem~\ref{thm.main_thm_noteBr}.

\begin{appendices}

\section{Transport of complex conjugation}\label{app.time_dep_conjugations_noteBr}

In this appendix, we prove Proposition~\ref{prop.time_dep_conjugations_noteBr}, stated below in a more abstract form.

\begin{lemma}\label{lemma.time_dep_conjugations_noteBr}
Let $\upphi = L + A$ be a bounded symplectomorphism on $\hi$, where $L$ and $A$ denote respectively the $\C$-linear and $\C$-antilinear parts of $\upphi$. Then, the $\C$-linear operator \[X = L^{\dagger} + A^{\dagger}\, \conjug\] satisfies the following properties:
\begin{enumerate}
    \item $X$ is invertible, 
    \item $\conjug_X \defegal X\, \conjug\, X^{-1}$ defines a complex conjugation on $\hi$, 
    \item $(\1 + \conjug)\, \upphi = X^{\dagger}\, (\1 + \conjug_X)$.
\end{enumerate}
\end{lemma}
\begin{proof}
\hspace{0mm}
\begin{enumerate}
    \item The two identities 
    \begin{equation}\label{identity_symplecto_noteBr}
        L\, L^\dagger - A\, A^\dagger = \1 \text{ and } L^\dagger\, L - A^\dagger\, A = \1\,,
    \end{equation}
    taken from \cite[Proposition~A.4]{Br12}, provide 
    \[\normhi{L z}^2 = \normhi{z}^2 + \normhi{A z}^2 \geq \normhi{z}^2 \text{ and } \normhi{L^\dagger z}^2 = \normhi{z}^2 + \normhi{A^\dagger z}^2 \geq \normhi{z}^2\] for all $z \in \hi$. The first inequality above shows that $L$ is injective with closed range, and the second one proves that $L^\dagger$ is injective, which amounts to the density of the range of $L$ in $\hi$. Consequently, $L$ is invertible and $\norm{L^{-1}}_{\mathcal{L}(\hi)} \leq 1$ (this result is stated in \cite[Proposition~A.5]{Br12}). Then, applying the first relation above to $z = L^{-1}\, w \in \hi$ provides 
    \[\normhi{w}^2 = \normhi{L^{-1}\, w}^2 + \normhi{A\, L^{-1}\, w}^2 \geq \norm{L}^{-2}\, \normhi{w}^2 + \normhi{A\, L^{-1}\, w}^2\] 
    for all $w \in \hi$, whence $\norm{A\, L^{-1}}^2 \leq 1 - \norm{L}^{-2} < 1$. Hence, the operator 
    \[X^\dagger = L + \conjug\, A = \left(\1 + \conjug\, A\, L^{-1}\right)\, L\] is invertible, and its inverse can be expressed as a Neumann series. The same holds for its adjoint $X$.
    \item The map $\conjug_X$ is $\C$-antilinear by $\C$-linearity of $X$, and satisfies $\conjug_X^2 = \1$. Hence, it suffices to prove that we have 
    \begin{equation}\label{isometric_cx_noteBr}
       \braket{z_1}{\conjug_X\, z_2} = \braket{z_2}{\conjug_X\, z_1} \text{ for all } z_1,z_2 \in \hi\,, 
    \end{equation}
from which it follows that $\conjug_X$ is isometric. Using the identity $\braket{z_1}{\conjug\, z_2} = \braket{z_2}{\conjug\, z_1}$, we obtain 
\[\braket{z_1}{\conjug_X\, z_2} = \braket{z_2}{(X^{-1})^\dagger\, \conjug\, X^\dagger\, z_1}\,.\] Hence, \eqref{isometric_cx_noteBr} holds if and only if $(X^{-1})^\dagger\, \conjug\, X^\dagger = \conjug_X = X\, \conjug\, X^{-1}$, which amounts to 
\begin{equation}\label{XdaggerX_commutes_with_c_noteBr}
    \comm{X^\dagger\, X}{\conjug} = 0\,.
\end{equation}
Let us prove the latter identity: using \eqref{identity_symplecto_noteBr} as well as $A\, L^{\dagger} = L\, A^{\dagger}$ (see, again, \cite[Proposition~A.4]{Br12}), we have
    \begin{align*}
        X^{\dagger}\, X = &L\, L^{\dagger} + \conjug\, A\, A^{\dagger}\, \conjug + \conjug\, A\, L^{\dagger} + L\, A^{\dagger}\, \conjug\\
        = &\1 + A\, A^{\dagger} + \conjug\, A\, A^{\dagger}\, \conjug + \conjug\, A\, L^{\dagger} + A\, L^{\dagger}\, \conjug\,,
    \end{align*}
    which yields $\conjug\, X^{\dagger}\, X\, \conjug = X^{\dagger}\, X$ and concludes the proof of \eqref{XdaggerX_commutes_with_c_noteBr}.
\item Using the decomposition $X^\dagger = L + \conjug\, A$, we obtain that $(\1 + \conjug)\, \upphi = (\1 + \conjug)\, X^\dagger$. Furthermore, we have 
\begin{align*}
    X^\dagger\, (\1 + \conjug_X) = &X^\dagger + X^\dagger\, X\, \conjug\, X^{-1}\\
    = &X^\dagger + \conjug\, X^\dagger\\
    = &(\1 + \conjug)\, X^\dagger\,,
\end{align*}
where we have used the commutation identity \eqref{XdaggerX_commutes_with_c_noteBr}. This completes the proof.
\end{enumerate}
\end{proof}

\section{Norm estimates for the quadratic propagator}\label{app.proof_crucial_estimate_noteBr}

The aim of this appendix is to provide a detailed proof of Proposition~\ref{prop.inegalite_cruciale_reste_propag_CS_QFT_noteBr}.

\bigskip

We start by establishing the existence of the unitary propagator for the quantum evolution as well as useful norm estimates, as announced in Subsection~\ref{subsect.op_Lambda_t_noteBr}. The following proof closely follows \cite[Subsection~4.2]{AmZe14}. We will use \cite[Corollary~C.4]{AmBr12}, stated below in a weaker version that fits our needs. 
	\begin{theorem}[\,{\cite[Corollary~C.4]{AmBr12}}]\label{thm.quadratic_dynamics_Ammari_Breteaux_noteBr}
		Let $I \subset \R$ be a compact interval containing $0$, and let ${k \in \N^*}$ be fixed. Let $(\mathfrak{Q}_t)_{t \in I}$ be a family of symmetric bounded operators in the space $\mathcal{L}\left(D\left(\nbop^{k/2}\right),D\left(\nbop^{k/2}\right)^*\right)$. Here, we set $\norm{\bullet}_{D\left(\nbop^{k/2}\right)} = \norm{(\nbop+1)^{k/2}\,\bullet}_{\focks}$. We assume that 
		\begin{enumerate}
			\item the map $t \mapsto \mathfrak{Q}_t \in \mathcal{L}\left(D\left(\nbop^{k/2}\right),D\left(\nbop^{k/2}\right)^*\right)$ is norm-continuous, 
			\item there exists a continuous function $f_k : I \to \R_+$ such that, for all $\eta,\psi \in \focksalg$ and $t \in I$, we have 
			\begin{equation}\label{estimee_commut_corollary_C4_Ammari_Breteaux_noteBr}
				\abs{\braket{(\nbop+1)^k\, \eta}{\mathfrak{Q}_t\, \psi} - \braket{\mathfrak{Q}_t\, \eta}{(\nbop+1)^k\, \psi}} \leq f_k(t)\, \norm{\eta}_{D\left(\nbop^{k/2}\right)}\, \norm{\psi}_{D\left(\nbop^{k/2}\right)}\,.
			\end{equation}
		\end{enumerate}
		Then, the Cauchy problem \begin{equation}\label{abstract_Cauchy_problem_quadratic_dynamics_noteBr}
		\left\{\begin{array}{l}
			\rmi\, \partial_t \psi_t = \mathfrak{Q}_t\, \psi_t\\
			\eval{\psi_t}_{t=0} = \psi_0 \in D\left(\nbop^{k/2}\right)
		\end{array}\right.
	\end{equation} admits a unique unitary propagator $(\tilde{U}_t)_{t\, \in \, I}$. 
	\end{theorem}

We wish to apply the previous theorem with $\mathfrak{Q}_t = \displaystyle\frac{1}{\eps}\, \Wick[\eps]{\tilde{Q}_t} = \Wick[1]{\tilde{Q}_t}$, where $\tilde{Q}_t$ is the quadratic symbol \eqref{def_quadratic_symbol_interaction_RPZ_noteBr}. These operators are symmetric since $\tilde{Q}_t$ is real-valued. Let us now verify that the assumptions of Theorem~\ref{thm.quadratic_dynamics_Ammari_Breteaux_noteBr} are satisfied:
\begin{enumerate}
    \item Writing \[\tilde{Q}_t(z) = \frac{1}{2\, \rmi}\, (b_{2,0}(t,z)-b_{0,2}(t,z))\] with $b_{2,0}(t,z) = \braket{\tilde{\mathfrak{b}}_t}{\tensn{z}{2}} \in \wickpol[2,0]$ and $b_{0,2}(t,z) = \braket{\tensn{z}{2}}{\tilde{\mathfrak{b}}_t} \in \wickpol[0,2]$, we have, for all $\psi \in \focksalg$, $t \in \R$ and $n \in \N$,
\begin{equation*}
    \left(\Wick[1]{\tilde{Q}_t}\, \psi\right)^{(n)} = \frac{1}{2\, \rmi}\, \Big(\Wick[1]{b_{2,0}(t)}\, \psi^{(n+2)} - \Wick[1]{b_{0,2}(t)}\, \psi^{(n-2)}\, \1_{n \geq 2}\Big)\,.
\end{equation*}
Hence, for all $k \in \N^*$ and for all $t,s \in \R$, using the definition of polynomial Wick quantization \eqref{quantif_Wick_poly_tensym_noteBr} with $(p,q) = (2,0)$ and $(p,q) = (0,2)$, we obtain:
\begin{align*}
    \norm{\Wick[1]{\tilde{Q}_t-\tilde{Q}_s}\, \psi}_{D\left(\nbop^{k/2}\right)^*}^2 \leq &\frac{1}{2}\, \sum_{n \geq 0} (n+1)^{-k}\, \norm{\Wick[1]{b_{2,0}(t)-b_{2,0}(s)}\, \psi^{(n+2)}}^2 \\
    &+ \frac{1}{2}\, \sum_{n \geq 2} (n+1)^{-k}\, \norm{\Wick[1]{b_{0,2}(t)-b_{0,2}(s)}\, \psi^{(n-2)}}^2\\
    \leq &\frac{1}{2}\, \sum_{n \geq 0} (n+1)^{-k}\, (n+1)\, (n+2)\, \norm{\tilde{\mathfrak{b}}_t-\tilde{\mathfrak{b}}_s}_{\tensn{\hi}{2}}^2\, \norm{\psi^{(n+2)}}^2 \\
    &+ \frac{1}{2}\, \sum_{n \geq 2} (n+1)^{-k}\, n\, (n-1)\, \norm{\tilde{\mathfrak{b}}_t-\tilde{\mathfrak{b}}_s}_{\tensn{\hi}{2}}^2\, \norm{\psi^{(n-2)}}^2\\
    \leq &\frac{1}{2}\, \norm{\tilde{\mathfrak{b}}_t-\tilde{\mathfrak{b}}_s}_{\tensn{\hi}{2}}^2\, \Big(\sum_{n \geq 0} (n+3)^k\, \norm{\psi^{(n+2)}}^2 \\
    &\phantom{\frac{1}{2}\, \norm{\tilde{\mathfrak{b}}_t-\tilde{\mathfrak{b}}_s}_{\tensn{\hi}{2}}^2\, \Big(\sum_{n \geq 2}}+ \sum_{n \geq 2} (n-1)^k\, \norm{\psi^{(n-2)}}^2\Big)\\
    \leq &\norm{\tilde{\mathfrak{b}}_t-\tilde{\mathfrak{b}}_s}_{\tensn{\hi}{2}}^2\, \norm{\psi}_{D\left(\nbop^{k/2}\right)}^2\,.
\end{align*}
Since $\tilde{\mathfrak{b}}_t$ is continuous by continuity of $\mathfrak{b}_t$ and strong continuity of $u_\mathfrak{a}(t)$, this proves that the map \[t \mapsto \mathfrak{Q}_t \in \mathcal{L}\left(D\left(\nbop^{k/2}\right),D\left(\nbop^{k/2}\right)^*\right)\] is norm-continuous.
\item We now establish an estimate of the form \eqref{estimee_commut_corollary_C4_Ammari_Breteaux_noteBr}. Let $\psi, \eta \in \focksalg$ and $t \in \R$. Using the formula 
		\[
		\left[(\nbop+1)^k, \Wick[\eps]{b}\right] \psi^{(n)} = \left((n+q-p+1)^k-(n+1)^k\right)\Wick[\eps]{b}\psi^{(n)}
		\] for $b \in \wickpol[p,q]$ and $\psi^{(n)} \in \tensymn{\hi}{n}$, an explicit computation shows that
		\begin{multline*}
		 \braket{(\nbop+1)^k\, \eta}{\Wick[1]{\tilde{Q}_t}\psi} - \braket{\Wick[1]{\tilde{Q}_t} \eta}{(\nbop+1)^k\, \psi}\\ = \braket{\eta}{\comm{(\nbop+1)^k}{\Wick[1]{\tilde{Q}_t}}\, \psi}
		\end{multline*}
		is equal to 
		\begin{multline*}
		    -\frac{1}{2\, \rmi}\, \sum_{n=0}^\infty \big((n+3)^k-(n+1)^k\big) \Big(\braket{\eta^{(n+2)}}{\Wick[1]{b_{0,2}(t)}\psi^{(n)}} \\+\braket{\eta^{(n)}}{\Wick[1]{b_{2,0}(t)}\psi^{(n+2)}}\Big)\,.
		\end{multline*}
		Since $(n+3)^k - (n+1)^k \leq 2\, k\, (n+3)^{k-1}$, we obtain 
		\begin{multline*}
			\abs{\braket{(\nbop+1)^k\, \eta}{\Wick[1]{\tilde{Q}_t}\psi} - \braket{\Wick[1]{\tilde{Q}_t} \eta}{(\nbop+1)^k\, \psi}}\\ \leq k\, \norm{\tilde{\mathfrak{b}}_t}_{\tensn{\hi}{2}} \sum_{n=0}^\infty (n+3)^{k-1} \sqrt{(n+1)(n+2)} \left(\norm{\eta^{(n+2)}}\norm{\psi^{(n)}} + \norm{\eta^{(n)}}\norm{\psi^{(n+2)}}\right)\,.
		\end{multline*}
		After dominating $(n+3)^{k-1} \sqrt{(n+1)(n+2)}$ by $3^{\frac{k-1}{2}}\, (n+1)^{k/2}\, (n+3)^{k/2}$, we can apply the Cauchy--Schwarz inequality, which provides   
		\begin{multline}\label{estimate_forme_quad_propag_quad_Pphi2}
			\abs{\braket{(\nbop+1)^k\, \eta}{\mathfrak{Q}_t\,\psi} -\braket{\mathfrak{Q}_t\,\eta}{(\nbop+1)^k\,\psi}}\\ 
			\leq 2\, k\, 3^{\frac{k-1}{2}} \norm{\tilde{\mathfrak{b}}_t}_{\tensn{\hi}{2}} \norm{\eta}_{D\left(\nbop^{k/2}\right)}\norm{\psi}_{D\left(\nbop^{k/2}\right)}.
		\end{multline}
\end{enumerate}
Finally, applying Theorem~\ref{thm.quadratic_dynamics_Ammari_Breteaux_noteBr} on each compact interval and using a gluing argument, we obtain that the quantum evolution problem \eqref{quantum_evolution_interaction_RPZ_noteBr} admits a unique unitary propagator $\left(\tilde{U}_t\right)_{t \in \R}$.

\bigskip

Let us now establish some norm estimates for $\tilde{U}_t$ in the Hilbert rigging: let $\psi \in \focksalg$ be fixed, and let $f(t) = \norm{\tilde{U}_t\, \psi}_{D\left(\nbop^{k/2}\right)}^2$. Applying Theorem~\ref{thm.quadratic_dynamics_Ammari_Breteaux_noteBr} at level $k+2$, we see that $\tilde{U}_t\, \psi \in D\left(\nbop^{\frac{k+2}{2}}\right)$, continuously in $t$. Furthermore, since $\mathfrak{Q}_t$ maps $D\left(\nbop^{\frac{k+2}{2}}\right)$ into $D\left(\nbop^{k/2}\right)$ norm-continuously in $t$, we obtain that $t \mapsto \tilde{U}_t\, \psi$ is $C^1$ in $D\left(\nbop^{k/2}\right)$. Then, the previous computation provides 
\begin{equation*}
    \abs{f'(t)} \leq 2\, k\, \norm{\tilde{\mathfrak{b}}_t}_{\tensn{\hi}{2}} \sum_{n=0}^\infty (n+3)^k\, \norm{(\tilde{U}_t\, \psi)^{(n+2)}}_{\tensn{\hi}{n+2}}\norm{(\tilde{U}_t\, \psi)^{(n)}}_{\tensn{\hi}{n}}\,.
\end{equation*}
Let us fix $\lambda_0 > 1$. There exists $n_0(\lambda_0) \in \N$ such that $(n+3)^k \leq \lambda_0^k\, (n+1)^k$ for all $n \geq n_0(\lambda_0)$. Hence, separating the right-hand side sum and applying the Cauchy--Schwarz inequality, we obtain 
\begin{equation*}
    \abs{f'(t)} \leq 2\, k\, c_0^k\, \norm{\tilde{\mathfrak{b}}_t}_{\tensn{\hi}{2}}\, \norm{\psi}_{\focks}^2 + 2\, k\, \lambda_0^k\, \norm{\tilde{\mathfrak{b}}_t}_{\tensn{\hi}{2}}\, f(t)\,,
\end{equation*}
where $c_0 = n_0(\lambda_0)+3$ is $k$-independent. Since $f(0) = \norm{\psi}_{D\left(\nbop^{k/2}\right)}^2$, Grönwall's lemma provides the following result, inspired by \cite[Proposition~4.5]{AmZe14}:
\begin{prop}\label{prop.prop_4.5_AmZe_noteBr}
For all $\lambda_0 > 1$, there exists $c_0 > 0$ such that we have, for all $k \in \N^*$, $t \in \R$ and $\psi \in \focksalg$,
\begin{multline}
    \norm{(\nbop+1)^{k/2}\, \tilde{U}_t\, \psi}_{\focks}^2\\ \leq \left(\norm{(\nbop+1)^{k/2}\, \psi}_{\focks}^2 + 2\, k\, c_0^k\, \abs{\int_0^t \norm{\tilde{\mathfrak{b}}_\tau}_{\tensn{\hi}{2}}\, \dd{\tau}}\, \norm{\psi}_{\focks}^2\right)\, \e^{2\, k\, \lambda_0^k\, \abs{\int_0^t \norm{\tilde{\mathfrak{b}}_\tau}_{\tensn{\hi}{2}}\, \dd{\tau}}}\,.
\end{multline}
This estimate can be extended by density to $\psi \in D\left(\nbop^{k/2}\right)$, using the unitarity of $\tilde{U}_t$ and the closedness of $(\nbop+1)^{k/2}$.
\end{prop}

\bigskip

We can now focus on the proof of Proposition~\ref{prop.inegalite_cruciale_reste_propag_CS_QFT_noteBr}, following \cite[Lemma~4.9]{AmZe14}. Let \[S(x) = \displaystyle\sum_{k = 0}^\infty \alpha_k\, x^k\] be an entire function with $\alpha_k > 0$, and let $\lambda_1 > 8\, \e$ be fixed (take for example ${\lambda_1 = 22}$, so that the dependence on $\lambda_1$ can be dropped in what follows). By \cite[Proposition~3.17]{AmZe14}, there exists $C > 0$ such that we have, for all $V \in \focks$ and $0 < \eps \leq 1/3$, 
\begin{multline}
    \norm{\Wick[\eps]{F_V}\, \Big(S(\nbop+1)\Big)^{-1/2}}_{\mathcal{L}(\focks)} \leq C\, \Bigg(\norm{\liftop{\sqrt{\eps}}\, V}_{\focks} \\+ \left(\sum_{k = 0}^\infty \frac{(\lambda_1\, \eps)^k}{\alpha_{k+2}}\, \norm{V^{(k)}}_{\tensn{\hi}{k}}^2\right)^{1/2}\Bigg)\,,
\end{multline}
whenever the right-hand side is finite. In particular, if $V \in D\left(\e^{\gamma\, \liftop{\lambda}}\right)$ for some $\gamma > 0$ and $\lambda > 1$ (see Assumption~\ref{hyp.potentiel_petit_noteBr}), choosing $\alpha_k = \e^{-\gamma_0 \lambda^k}$ with $0 < \gamma_0\, \lambda^2 < 2 \gamma$, we obtain:
\begin{align*}
    \norm{\Wick[\eps]{F_V}\, \Big(S(\nbop+1)\Big)^{-1/2}}_{\mathcal{L}(\focks)}^2
    &\leq C\, \left(\norm{\liftop{\sqrt{\eps}}\, V}_{\focks}
    + \norm{\e^{\gamma \liftop{\lambda}}\liftop{\sqrt{\eps}}V}_{\focks}\right)^2\\
    &\leq C\, \norm{\e^{\gamma \liftop{\lambda}}\liftop{\sqrt{\eps}}V}_{\focks}^2\,,
\end{align*}
for some $\eps$-independent $C > 0$ that may change from line to line
and depends only on $\gamma_0$, $\gamma$ and $\lambda$ (in particular, not directly on $V$). Furthermore, for $\lambda > \lambda_0 > 1$ fixed, summing the result of Proposition~\ref{prop.prop_4.5_AmZe_noteBr} over $k$, we have, for all $\psi \in \focksalg$,
\begin{equation*}
    \norm{\Big(S(\nbop+1)\Big)^{1/2}\, \tilde{U}_t\, \psi}_{\focks}^2 \leq \norm{\Big(\mathrm{g}_t(\nbop+1)\Big)^{1/2}\, \psi}_{\focks}^2 + 2\, c_0\, \mathrm{g}_t'(c_0)\,\abs{\int_0^t \norm{\tilde{\mathfrak{b}}_\tau}_{\tensn{\hi}{2}} \dd{\tau}}\cdot \norm{\psi}_{\focks}^2\,,
\end{equation*}
where we have set \[\mathrm{g}_t(x) = \sum_{k=0}^\infty \alpha_k(t)\, x^k\,,\quad \alpha_k(t) = \alpha_k\, \e^{2\, k\, \lambda_0^k\, \abs{\int_0^t \norm{\tilde{\mathfrak{b}}_\tau}_{\tensn{\hi}{2}}\, \dd{\tau}}}\]
and where the right-hand side is finite since $\psi \in \focksalg$. Finally, the two previous estimates provide 
\begin{multline*}
			\norm{\Wick[\eps]{F_V} \, \tilde{U}_t \, \psi}_{\focks} \leq C \norm{\e^{\gamma \liftop{\lambda}} \liftop{\sqrt{\eps}} V}_{\focks}\times\\
			\left(\norm{\Big(\mathrm{g}_t(\nbop+1)\Big)^{1/2}\, \psi}_{\focks}^2 + 2\, c_0 \, \mathrm{g}_t'(c_0) \abs{\int_0^t \norm{\tilde{\mathfrak{b}}_\tau}_{\tensn{\hi}{2}} \dd{\tau}} \norm{\psi}_{\focks}^2\right)^{1/2}\,,
		\end{multline*}
which is exactly Proposition~\ref{prop.inegalite_cruciale_reste_propag_CS_QFT_noteBr} with $U_t$ and $\mathfrak{b}_t$ replaced by $\tilde{U}_t$ and $\tilde{\mathfrak{b}}_t$. We conclude the proof of Proposition~\ref{prop.inegalite_cruciale_reste_propag_CS_QFT_noteBr} by recalling that $U_t = \liftop{u_\mathfrak{a}(t)}\, \tilde{U}_t$ and $\mathfrak{b}_t = \liftop{u_\mathfrak{a}(t)}\, \tilde{\mathfrak{b}}_t$, where the operator $\liftop{u_\mathfrak{a}(t)}$ is unitary and commutes with the number operator $\nbop$, thus with $\Big(S(\nbop+1)\Big)^{1/2}$. Consequently, we have $\norm{\tilde{\mathfrak{b}}_t}_{\tensn{\hi}{2}} = \norm{\mathfrak{b}_t}_{\tensn{\hi}{2}}$, and $\tilde{U}_t$ can be replaced by $U_t$ in the estimate of $\norm{\Big(S(\nbop+1)\Big)^{1/2}\, \tilde{U}_t\, \psi}_{\focks}$.

\end{appendices}

\section*{Acknowledgments}

The author wishes to express his gratitude to Sébastien Breteaux and Jérémy Faupin for a very enlightening discussion held at Institut Élie Cartan de Lorraine (Metz, France) in November 2025. The author is also grateful to the CUBO referees for their careful review of the manuscript and for their constructive comments and valuable suggestions.

\section*{Statements and Declarations}

The author declares that there is no conflict of interest.

\section*{Data Availability Statement}

No data sets were generated or analyzed during this study.

\printbibliography
	
\end{document}